\documentclass[12pt]{article}
 \usepackage{sprocl_mod}
 \usepackage{rotating}
 \usepackage{epsfig}
 \usepackage{graphicx}
\newcommand{\zf}{{\tt{ZFITTER}}}
\newcommand{\flm}{\mu}
\newcommand{\mz}{M_{_Z}}
\newcommand{\afba}[1]{A^{#1}_{_{\rm FB}}}

\def\be{\begin{equation}}
\def\ee{\end{equation}}
\def\ba{\begin{eqnarray}}
\def\ea{\end{eqnarray}}
\def\nl{\nonumber\\}
\begin{document}
\thispagestyle{empty}
\begin{flushleft}
{\tt
DESY 99-037
\\
August 1999
\\
hep-ph/9908289
}
\end{flushleft}

\bigskip

\title{%
PREDICTIONS 
OF {\tt \large ZFITTER} v.6
FOR 
\\
FERMION-PAIR PRODUCTION WITH ACOLLINEARITY CUT
\footnote{Based on talks presented at
ECFA/DESY Linear  
                  Collider Project meetings held at Frascati,
		  Nov 1998, 
and 
Oxford, March 1999, 
at the LEP~2 Mini-Workshop held
		  at CERN, {M}arch 1999, 
and at the Workshop for a
		  Worldwide Study on Physics and Experiments with
		  Future Linear $e^+e^-$ Colliders held at
		  Sitges/Barcelona, April 1999.
}}
\bigskip

\author{%
P. CHRISTOVA\footnote{Supported by Bulgarian Foundation for
Scientific Research with grant $\Phi$--620/1996.}  
}
\address{%
Faculty of Physics, Bishop Konstantin Preslavsky Univ., 
\\ Shoumen, Bulgaria
\\
and
\\
Lab. for Nuclear Problems, Joint Institute for Nuclear Research, 
\\
Dubna, Russia
\\ E-mail: penchris@nusun.jinr.ru
} 
\author{M. JACK, S. RIEMANN, T. RIEMANN
}
\address{DESY Zeuthen, Platanenallee 6,
\\ D-15738 Zeuthen, Germany
\\ E-mails: jack@ifh.de, riemanns@ifh.de, riemann@ifh.de
}
\maketitle

\bigskip

\abstracts{
{\tt ZFITTER} is a Fortran package for the description of fermion-pair
production in $e^+e^-$-annihilation.
We report on results of a rederivation of the complete set of 
analytical $O(\alpha)$ formulae for the treatment of 
photonic corrections to the
total cross-section and the integrated forward-backward asymmetry
with combined cuts on acollinearity angle, acceptance angle, and
minimal energy of the fermions.
Numerically, 
the following changes result in {\tt ZFITTER}
v. 6.11 compared to {\tt ZFITTER} v. 5.20/21:
(i) at the $Z$ resonance -- numerical changes are negligible;
(ii) at LEP~1 energies off-resonance -- corrections amount to at most few per
mil;  
(iii) at LEP~2 energies -- corrections amount to one per cent or less.
Thus, the predictions for   
LEP/SLC data remain unchanged within the actual errors.
}

\clearpage

\section{\Large \bf Introduction
}
The Fortran package \zf\  
\cite{DESY99070,Bardin:1992jc2,Bardin:1991de,Bardin:1991fu,Bardin:1989di,%
Christova:1999cc}
is regularly used for the analysis of LEP data since 1989 for the
reaction:
\ba
\label{reaction}
e^+(k_2) + e^-(k_1) \to {\bar f}(p_2) + f(p_1) + (n\gamma)(p)
.
\ea
Since the systematic description of \zf\  v.4.5 (19 April 1992) 
\cite{zfitter:v4.5a} 
a series of improved versions has been released
and a nearly complete collection of them may be found in the world wide web 
\cite{afs-Bardin,home-Riemann2}.
Most important updates of recent years are related to the treatment of 
higher-order corrections.
In 1995, this was documented systematically in
\cite{Bardin:1995aa}.
Later improvements are described in \cite{DESY99070}.
The \zf\ versions 5.20/21 
\cite{zfitter:v5.2021m} 
were quite recently followed by versions v.6.nm, beginning with v.6.04/06
\cite{zfitter:v6.0406m}. 

Traditionally we aimed at an accuracy of \zf\ at LEP~1 energies of the order of
0.5 \%. 
The successful running of LEP~1 together with the precise
knowledge of 
the beam energy however makes an even higher precision necessary
\cite{Christova:1998tc}:
We expect for the final measurements relative errors for total
cross-sections and absolute errors for asymmetries of up to 0.15\%
 at the $Z$ peak and of up to 0.5\%
 at $\sqrt{s} = M_Z~ \pm$ several GeV;
aiming from the theoretical side ideally at a tenth of these values
for the errors of single
corrections, we estimate limits of 0.015\% and 0.05\%, respectively
\footnote
{This might not even be sufficient if the Giga-Z option of the TESLA  
project will be realised (see e.g. \cite{Moenig:1999aa}), 
with a factor of 10 or 100 more $Z$ bosons produced than at LEP~1.
}. 
Similar claims may be found in \cite{Bardin:1999gt}.

\begin{figure}[tbhp] 
\begin{flushleft}
  \begin{tabular}{ll}
    \mbox{
 \epsfig{file=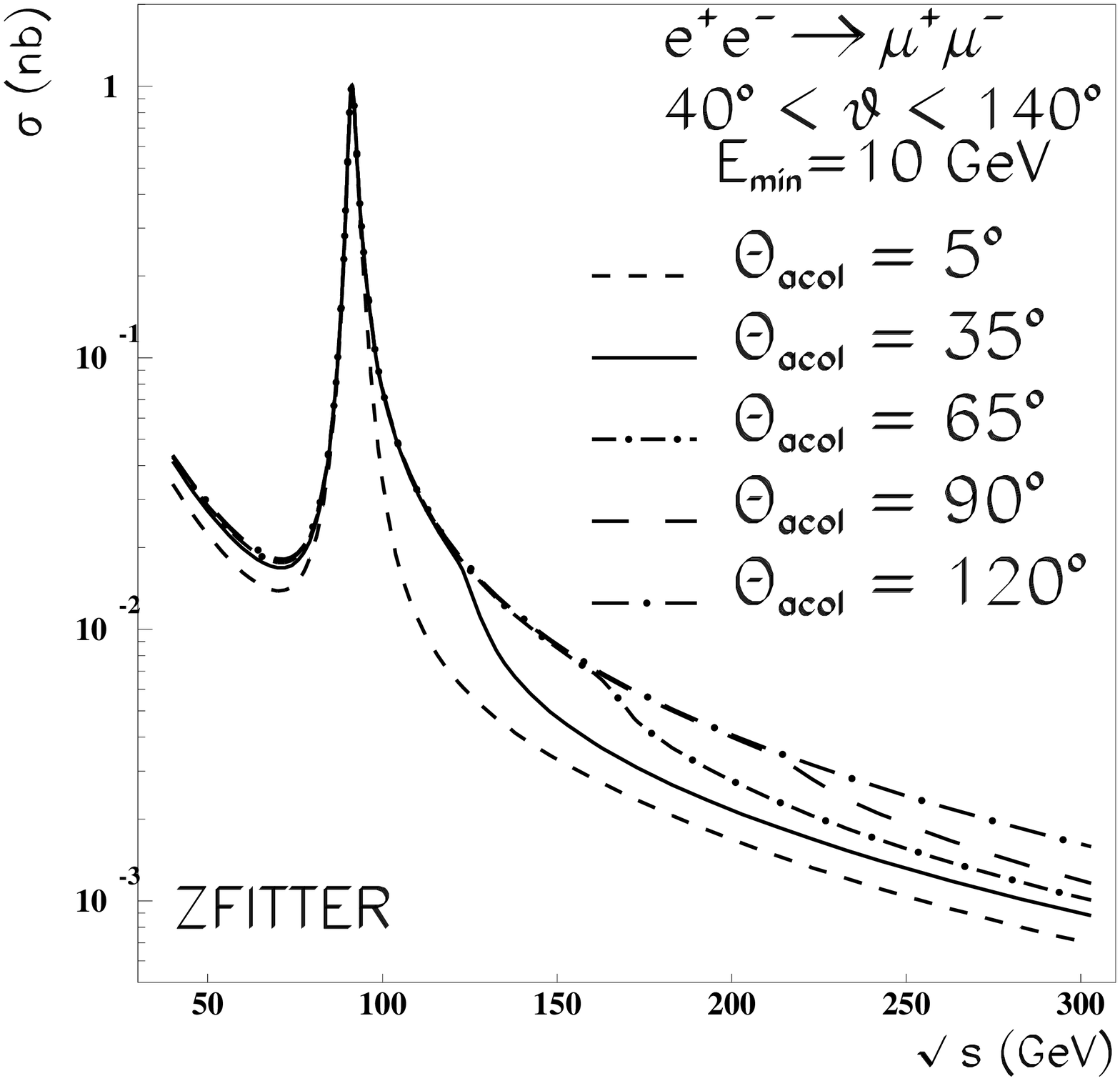,width=7.7cm}}
    &
\mbox{
 \epsfig{file=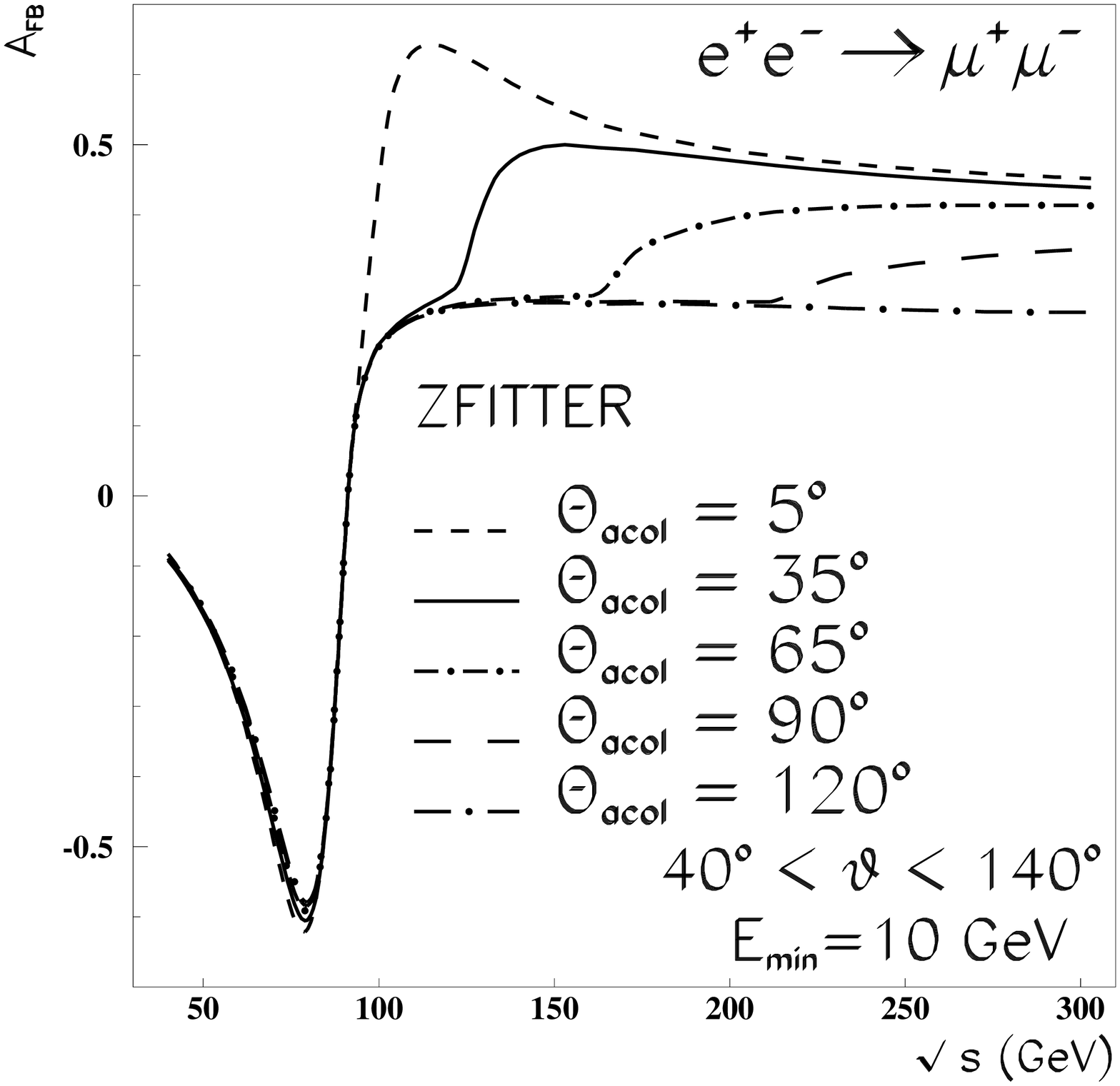,width=7.7cm}}
  \end{tabular}
\end{flushleft}
\caption[]{
Cross-sections and
forward-backward asymmetries for muon-pair production
with different acollinearity cuts.
\label{sigma3} 
} 
\end{figure}

\begin{figure}[bhtp]
 \begin{center}
  \vspace*{-1.0cm}
  \hspace*{-1.0cm}
  \mbox{
  \epsfig{file=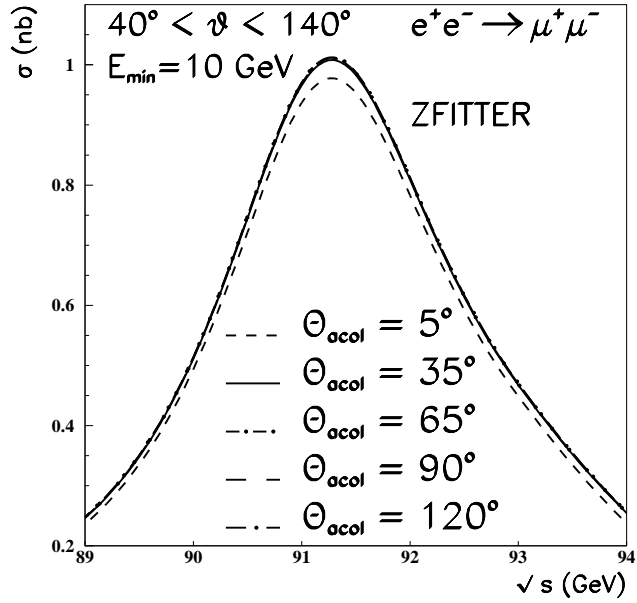,height=9.0cm}}
 \caption
 {
 \sf
 Muon-pair production cross-sections at LEP~1 energies with different
 acollinearity cuts.
 \label{sigma2}
 }
 \end{center}
 \end{figure}

First applications of \zf\ at energies above the $Z$ resonance have 
become relevant since data from LEP~1.5 and LEP~2 
are being analysed.
There is also rising interest in applications for the study of the physics
potential of a Linear Collider operating at $\sqrt{s} \geq 500$ GeV 
\cite{Accomando:1997wt}.
At the higher energies, again one generally expects final experimental 
accuracies of up to 0.8\% 
\cite{Christova:1998tc}.
So, one should aim at theoretical accuracies of single corrections of up to
0.1 \%. 
The excellent precision of \zf\ at the $Z$ peak, however,
does not automatically guarantee a sufficient accuracy at higher
energies, especially since the 
hard photonic
contributions, including higher-order corrections, are no longer suppressed; see
Figure 
\ref{sigma3} and compare to Figure \ref{sigma2}.

\begin{figure}[bhtp] 
\begin{center} 
 \vspace*{-5.0cm} 
 \hspace*{-4.5cm} 
\mbox{\epsfig{file=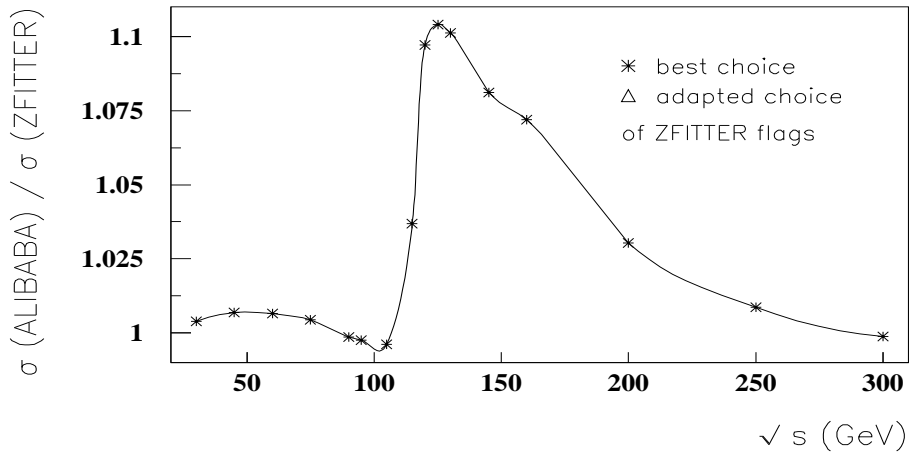,%
height=25.cm,%
width=20.cm,%
                   clip=%
} }
\vspace*{-13.cm} 
 
\caption{
\sf
Muon-pair production cross-section ratios 
{\tt ALIBABA} v.1 (1990) versus {\tt ZFITTER} v.4.5 (1992).
An acollinearity cut was applied with $\theta_{\rm acol}^{\max}=25^{\circ}$.
\label{compar1992} 
} 
\end{center} 
\end{figure}

In 1992, a comparison of {\tt ALIBABA} v.1 (1991)
\cite{Beenakker:1991mb}
and \zf\ v.4.5 (1992) showed
deviations between the predictions of the two programs of several per cent
\cite{Riemann:199200}; we show one of the plots of that study
in Figure \ref{compar1992}. 
These deviations were observed only above the $Z$ peak and only when
an acollinearity cut on the fermions was applied;
the agreement was much better without this cut.
We repeated the comparison in 1998 
with {\tt ALIBABA} v.2 (1991),
{\tt TOPAZ0} v.4.3 (1998) 
\cite{Montagna:1998kp,Passarino:199800},
and \zf\ v.5.14 (1998).
The outcome was basically unchanged compared to 1992 as may be seen in
Figure 3 of reference \cite{Christova:1998tc}.

It is well-known that deviations of up to several per cent may result from
different treatments of radiative corrections.
Recent studies 
\cite{Placzek:1999xc}
claim for the case of Bhabha scattering that an accuracy of 0.3\% for
$O(\alpha)$ corrections and of 1\% 
for the complete corrections has been reached at LEP~2 energies as long
as the radiative return to the $Z$ peak is prevented by cuts.
Similar conclusions were drawn in \cite{Montagna:1997jt} for
fermion-pair production.
The radiative return is prevented if $\sqrt{s'} > M_Z$, i.e. if $R=s'/s >
M_Z^2/s$.
Our figures contain predictions with an acollinearity cut of  
$\theta_{\rm acol}<10^{\circ},25^{\circ}$. 
This corresponds to an $s'$-cut of roughly $R  > R_{\theta_{\rm acol}} =
0.84,0.64 $ and $\sqrt{s} >$ 100 GeV, 114 GeV, respectively 
(see also Table \ref{rxivalues} in the Appendix).  
The resulting suppression of the radiative return at these and higher
energies can nicely be observed in Figure 
\ref{sigma3}.
But even if the radiative return is prevented, the influence of hard photonic
corrections will be much larger at higher energies than it is near the $Z$
resonance where hard bremsstrahlung is nearly
completely suppressed.      
Figure \ref{sigma2}
demonstrates that different portions of hard photon
emission lead to nearly identical cross-sections unless the region is
reached where even soft photon  emission is touched (lowest lying curve).

\bigskip

In this paper, we report on a recalculation of the 
photonic corrections with acolli\-near\-ity cut in the \zf\ approach.   
In Section \ref{phot-corr}, we explain changes of flags, subroutines,
kinematics, and numerics of  \zf. 
The basic differences to the simpler $s'$-cut are explained in
subsection \ref{sec-defi}.
Section \ref{summary} contains few comparisons with other programs and
Section \ref{summary2} a summary. 

\section{\Large \bf Photonic Corrections with Acollinearity Cuts
\label{phot-corr}   }
Perhaps we should remark here that the photonic  $O(\alpha)$ corrections in
\zf\ 
remained basically untouched since about 1989. 
For the $s'$-cut, \zf\ relies on duplicated analytical 
calculations \cite{Bardin:1987hva,Bardin:1989cw,Bardin:1991de,Bardin:1991fu}, 
and numerical comparisons showed the reliability of the
predictions at LEP energies; see e.g. \cite{Bardin:1999gt} for LEP~1 and
\cite{Boudjema:1996qg} for LEP~2. 
Concerning the acollinearity cut, the situation is different.
The corresponding part of \zf\ \cite{MBilenky:1989ab} was never
checked independently,  and is not documented\footnote
{ 
A collection of some formulae related to the initial-state
corrections (and its combined exponentiation with final-state radiation) for
the angular distribution may be found in 
\cite{Bilenkii:1989zg}.
}.
So, we started 
a complete revision of the $O(\alpha)$
contributions with acollinearity cut.
First results were published in 
\cite{Christova:1998tc} 
where we reported that some deviations from \zf\ were found for
initial-state radiation.
We also reported that these deviations could not explain the differences
observed when comparing with {\tt ALIBABA}, while the treatment of higher
order corrections evidently was of much higher influence in this respect. 
Slightly later
it was observed in \cite{Bardin:1999gt}
that the perfect agreement of many
predictions of \zf\ v.5.20 and {\tt TOPAZ0} v.4.3 at LEP~1 energies of
about typically 0.01\% 
could not be reproduced when an acollinearity cut was applied and the 
initial-final state interference was taken into account.
This second puzzle could be resolved by our recalculation; see section
\ref{interference}. 

By now we have a complete collection of the analytical formulae for the
$O(\alpha)$ corrections.
The corresponding Fortran package is {\tt acol.f}.
We merged package {\tt acol.f} 
with photonic corrections for the integrated
total cross-section and the integrated forward-backward asymmetry (with and
without acceptance cut) into \zf\ v.5.21, thus creating \zf\ v.6.04/06
\cite{zfitter:v6.0406m} onwards. 
The angular distribution will be available in v.6.2 onwards.
The remarkably compact expressions for the case that no angular acceptance cut
is applied are published \cite{Christova:1999cc}. 
A complete collection of the analytical expressions is in preparation
\cite{DESY:1999??}.

\subsection{\large \bf Flags and subroutines in \zf
\label{flags}
}
In \zf\ the calculation of cross sections and asymmetries is 
dealt with by subroutine {\tt ZCUT} which either calls subroutines
{\tt SCUT} (angular acceptance cut applied) or {\tt SFAST} (shorter
formulae; no angular acceptance cut).
In the latest releases, \zf\ v.5.20 up to \zf\ v.6.11, 
the following flag settings in subroutine {\tt ZUCUTS} are of 
relevance here:
\begin{itemize}
\item
{\tt ICUT} = --1: cut on the invariant mass of the fermion pair, $s'$
\item
{\tt ICUT} = 0: 
cuts on acollinearity and minimal energy of the fermions and on the acceptance
angle of one fermion
\item
{\tt ICUT} = 1:
cuts on $s'$ and on the acceptance angle $\vartheta$ of one fermion
\end{itemize}
Since \zf\ v.6.04, we introduced two new values of this flag:
\begin{itemize}
 \item
{\tt ICUT} = 2:
new coding  of case corresponding to {\tt ICUT} = 0, but without acceptance
cut 
\cite{Christova:1999cc};
\item
{\tt ICUT} = 3:
 new coding of case corresponding to {\tt ICUT} = 0 \cite{DESY:1999??}.
\end{itemize}
In order to maintain compatibility with earlier releases, the branch
{\tt ICUT}=0 is retained; it
calculates with the old coding of the final-state corrections
(flag {\tt IFUNFIN} = 0),
while all other branches use the newly corrected final-state contributions
(flag {\tt IFUNFIN} = 1).

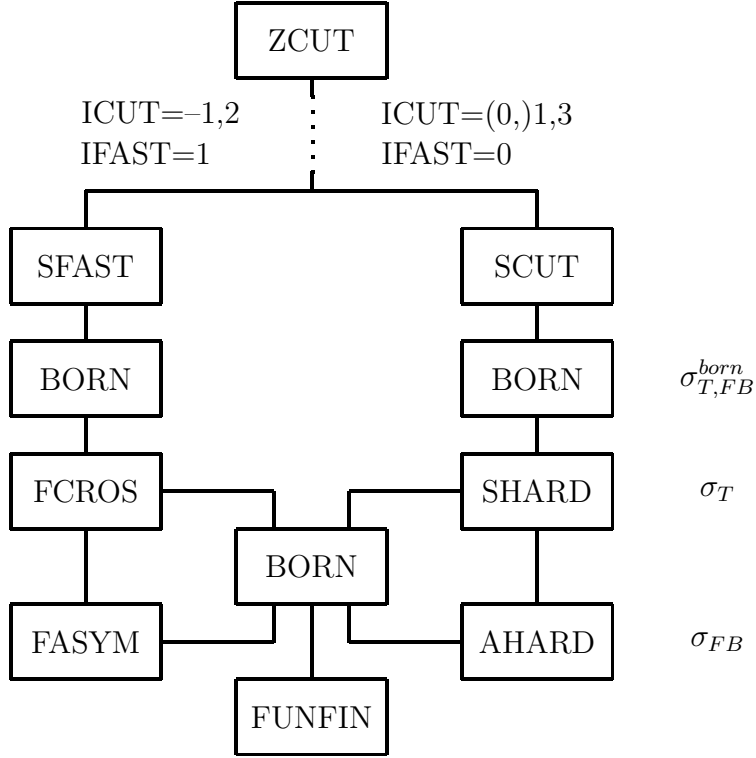
\begin{figure} 
\begin{center}
\vspace{2cm}

\unitlength=1.00mm
\special{em:linewidth 0.4pt}
\linethickness{0.4pt}
\begin{picture}(150.00,100.00)(-15,0)
\linethickness{1pt}
\put(16.00,040.00){\line(1,0){20.00}}
\put(16.00,030.00){\line(1,0){20.00}}
\put(36.00,040.00){\line(0,-1){10.00}}
\put(16.00,040.00){\line(0,-1){10.00}}
\put(26.00,035.00){\makebox(0,0)[cc]{FASYM}}
\put(76.00,040.00){\line(1,0){20.00}}
\put(76.00,030.00){\line(1,0){20.00}}
\put(76.00,040.00){\line(0,-1){10.00}}
\put(96.00,040.00){\line(0,-1){10.00}}
\put(86.00,035.00){\makebox(0,0)[cc]{AHARD}}
\put(16.00,060.00){\line(1,0){20.00}}
\put(16.00,050.00){\line(1,0){20.00}}
\put(36.00,060.00){\line(0,-1){10.00}}
\put(16.00,060.00){\line(0,-1){10.00}}
\put(26.00,055.00){\makebox(0,0)[cc]{FCROS}}
\put(76.00,060.00){\line(1,0){20.00}}
\put(76.00,050.00){\line(1,0){20.00}}
\put(76.00,060.00){\line(0,-1){10.00}}
\put(96.00,060.00){\line(0,-1){10.00}}
\put(86.00,055.00){\makebox(0,0)[cc]{SHARD}}
\put(16.00,090.00){\line(1,0){20.00}}
\put(16.00,080.00){\line(1,0){20.00}}
\put(36.00,090.00){\line(0,-1){10.00}}
\put(16.00,090.00){\line(0,-1){10.00}}
\put(26.00,085.00){\makebox(0,0)[cc]{SFAST}}
\put(76.00,090.00){\line(1,0){20.00}}
\put(76.00,080.00){\line(1,0){20.00}}
\put(76.00,090.00){\line(0,-1){10.00}}
\put(96.00,090.00){\line(0,-1){10.00}}
\put(86.00,085.00){\makebox(0,0)[cc]{SCUT}}
\put(46.00,050.00){\line(1,0){20.00}}
\put(46.00,040.00){\line(1,0){20.00}}
\put(46.00,050.00){\line(0,-1){10.00}}
\put(66.00,050.00){\line(0,-1){10.00}}
\put(56.00,045.00){\makebox(0,0)[cc]{BORN}}
\put(46.00,030.00){\line(1,0){20.00}}
\put(46.00,020.00){\line(1,0){20.00}}
\put(46.00,030.00){\line(0,-1){10.00}}
\put(66.00,030.00){\line(0,-1){10.00}}
\put(56.00,025.00){\makebox(0,0)[cc]{FUNFIN}}
\put(46.00,120.00){\line(1,0){20.00}}
\put(46.00,110.00){\line(1,0){20.00}}
\put(46.00,120.00){\line(0,-1){10.00}}
\put(66.00,120.00){\line(0,-1){10.00}}
\put(56.00,115.00){\makebox(0,0)[cc]{ZCUT}}
\put(56.00,110.00){\line(0,-1){2.50}}
\bezier{5}(56.00,107.50)(56.00,102.50)(56.00,097.50)
\put(56.00,097.50){\line(0,-1){2.50}}

\put(56.00,095.00){\line(-1,0){30.00}}
\put(56.00,095.00){\line(1,0){30.00}}

\put(26.00,095.00){\line(0,-1){5.00}}
\put(86.00,095.00){\line(0,-1){5.00}}

\put(26.00,050.00){\line(0,-1){10.00}}
\put(86.00,050.00){\line(0,-1){10.00}}
\put(56.00,040.00){\line(0,-1){10.00}}

\put(36.00,055.00){\line(1,0){15.00}}
\put(76.00,055.00){\line(-1,0){15.00}}
\put(36.00,035.00){\line(1,0){15.00}}
\put(76.00,035.00){\line(-1,0){15.00}}

\put(51.00,055.00){\line(0,-1){5.00}}
\put(61.00,055.00){\line(0,-1){5.00}}
\put(51.00,035.00){\line(0,1){5.00}}
\put(61.00,035.00){\line(0,1){5.00}}

\put(26.00,080.00){\line(0,-1){5.00}}
\put(86.00,080.00){\line(0,-1){5.00}}
\put(16.00,075.00){\line(0,-1){10.00}}
\put(76.00,075.00){\line(0,-1){10.00}}
\put(36.00,075.00){\line(0,-1){10.00}}
\put(96.00,075.00){\line(0,-1){10.00}}
\put(16.00,075.00){\line(1,0){20.00}}
\put(76.00,075.00){\line(1,0){20.00}}
\put(16.00,065.00){\line(1,0){20.00}}
\put(76.00,065.00){\line(1,0){20.00}}
\put(26.00,070.00){\makebox(0,0)[cc]{BORN}}
\put(86.00,070.00){\makebox(0,0)[cc]{BORN}}

\put(26.00,065.00){\line(0,-1){5.00}}
\put(86.00,065.00){\line(0,-1){5.00}}

\put(110.00,070.00){\makebox(0,0)[cc]{$\sigma^{born}_{T,FB}$}}
\put(110.00,055.00){\makebox(0,0)[cc]{$\sigma_T$}}
\put(110.00,035.00){\makebox(0,0)[cc]{$\sigma_{FB}$}}

\put(036.00,105.00){\makebox(0,0)[cc]{ICUT=--1,2}}
\put(034.00,100.00){\makebox(0,0)[cc]{IFAST=1}}
\put(078.00,105.00){\makebox(0,0)[cc]{ICUT=(0,)1,3}}
\put(074.00,100.00){\makebox(0,0)[cc]{IFAST=0}}

\linethickness{0.5pt}
\end{picture}

\vspace{-2cm}
\caption[Logical structure of {\tt ZCUT}]
{\centering{Logical structure of subroutine {\tt ZCUT}. 
}}
\label{structZCUT}
\end{center}
\end{figure}

All the corrections related to the acollinearity cut are called from
essentially only six subroutines/functions of \zf; see Figure 
\ref{structZCUT}.
The subroutines {\tt SHARD} and {\tt AHARD}, or respectively 
{\tt FCROS} and {\tt FASYM}, call the initial-state and initial-final 
state interference corrections, while  {\tt FUNFIN} calculates final-state
corrections (for the case of common exponentiation of initial- and
final-state soft photon corrections).

For a detailed description of the Fortran package \zf\ we refer to
\cite{DESY99070}. 

\subsection{\large \bf Photonic corrections with acollinearity cut:
Kinematics 
\label{sec-defi}
}
\subsubsection{The phase-space parameterisation}
\begin{figure}[tbhp]
  \begin{center}
\vspace*{-1.2cm}
  \mbox{%
  \epsfig{file=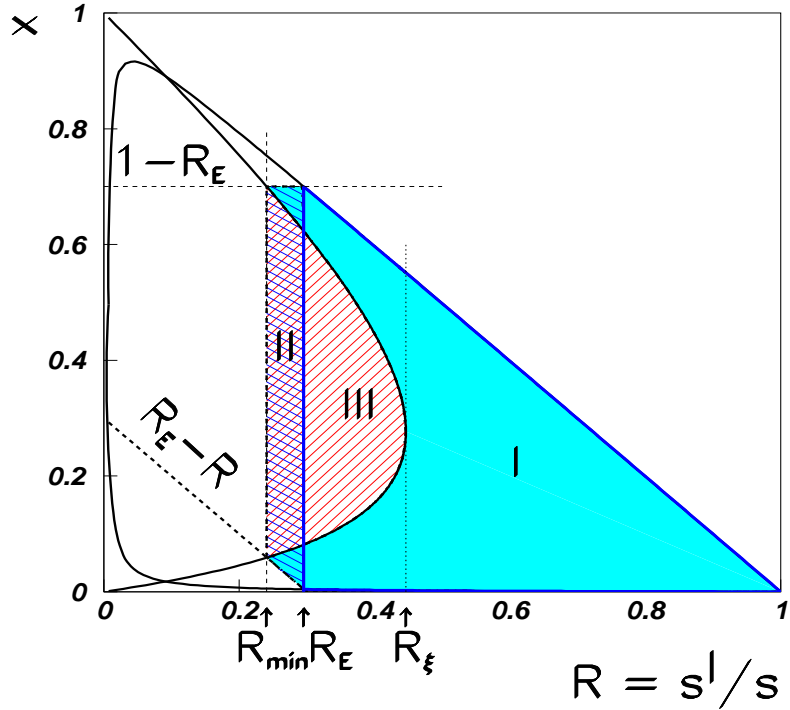,height=11cm,width=12cm}}
\caption{\label{dalitz} 
\sf
 Phase-space with cuts on acollinearity and energy of the fermions. 
}
  \end{center}
\end{figure}

The phase-space parameterisation
derived in \cite{Passarino:1982zp} is used.
A three-fold analytical integration of the squared matrix elements
had to be performed.
The last 
integration, that over $R=s'/s$, is then performed  numerically. 
The Dalitz plot given in Figure \ref{dalitz} may help to understand the 
relation between a kinematically simple $s'$-cut and a more involved
acollinearity cut.
The variable shown besides $R$ is $x$, the invariant mass of
(${\bar f}+\gamma$) in the cms. 
As Figure \ref{dalitz} shows, we have to determine the cross-sections 
in three phase-space regions with different boundary values of $x$ at
given $R$:  
\ba
\frac{d\sigma}{d\cos\vartheta} = \left[ \int_{\mathrm{I}} +
\int_{\mathrm{II}}   
- \int_{\mathrm{III}} \right]~ dR ~ dx 
\frac{d\sigma}{dR dx d\cos\vartheta} .
\label{sig}
\ea
Region I corresponds to the simple $s'$-cut.

The integration over $R$ extends from $R_{min}$ to 1,
\ba
\label{rmin}
R_{min} &=& R_E \left(1 - \frac{\sin^2(\theta_{\rm acol}^{\max}/2)}
{1-R_E\cos^2(\theta_{\rm acol}^{\max}/2)} \right) . 
\ea
The soft-photon corner of the phase space is at $R=1$.
Thus, the additional contributions related to the acollinearity cut are
exclusively due to hard photons.
The boundaries for the integration over $x$ are, for a given value of
$R$:
\ba
\label{ibound}
x_{max,min} &=& \frac{1}{2} (1-R)~ (1 \pm A),
\ea
where the parameter $A=A(R)$ depends in every region on only one of
the cuts applied:  
\ba
A_{\mathrm{I}} &=& \sqrt{1-\frac{R_{m}}{R}} \approx 1,
\label{A1}
\\
A_{\mathrm{II}} &=& \frac{1+R-2R_E}{1-R}, 
\label{A2}
\\
A_{\mathrm{III}} &=& \sqrt{1 - \frac{R(1-R_{\theta_{\rm
acol}})^2}{R_{\theta_{\rm acol}}(1-R)^2}},  
\label{A3}
\ea
with:
\begin{eqnarray}
  \label{eq:rs}
R_{m}  &=& \frac{4m_f^2}{s},
\\
  \label{eq:re}
R_E &=& \frac{2E_{min}}{\sqrt{s}}, 
\\
  \label{eq:rx}
R_{\theta_{\rm acol}} &=&
\frac{1-\sin(\theta_{\rm acol}^{\max}/2)}{1+\sin(\theta_{\rm
acol}^{\max}/2)}.     
\end{eqnarray}
Here, $m_f$ and $E_{min}$ are the final-state fermions' mass and a cut on
their individual energies in the cms.

\subsubsection{Mass singularities from initial-state radiation}
\label{sec:cases}
In the last section we saw that the Dalitz plot in Figure \ref{dalitz} 
is independent of the scattering angle $\cos\vartheta$.
Here we will show that the integrations in regions II and III are
nevertheless crucially influenced by   $\cos\vartheta$. 

The squared matrix elements contain, from initial-state radiation, 
the electron (positron) propagator, and these terms are
proportional to first and second powers of 
\ba
\frac{1}{Z_{1(2)}} &=& - \frac{1}{(k_{1(2)}-p)^2-m_e^2} 
    = \frac{1}{2k_{1(2)}p}
\nl &=& \frac{1}{A_{1(2)} \pm B \cos\varphi_\gamma},
\label{z1}
\ea
with
\ba
A_{1(2)}&=& \frac{s}{2}  (1-R)(1 \pm \beta_0 \cos\vartheta 
\cos\theta_{\gamma} ),
\\
B &=& \frac{s}{2}  (1-R)\beta_0 \sin \vartheta \sin \theta_{\gamma},
\ea
and
\ba
\beta_0 &=& \sqrt{1- 4m_e^2/s},
\\
\cos\theta_{\gamma}&=& \frac{\lambda_1 - \lambda_2 -\lambda_p }
{2\sqrt{\lambda_2\lambda_p}},
\\
\sqrt{\lambda_1} &=& (1-x)s,
\\
\sqrt{\lambda_2} &=& (x+R)s, 
\\
\sqrt{\lambda_p} &=& (1-R)s.
\ea
For the calculation of the initial-state corrections 
we may neglect final-state mass effects. 

The first analytical integration is over $\varphi_\gamma$, the
photon production angle in the ($f+\gamma$) rest system
\cite{Passarino:1982zp}:
\ba
\frac{d\sigma}{dR dx d\cos\vartheta} &=& 
  \int_{0}^{2\pi} d\varphi_\gamma \frac{d\sigma}{dR dx d\cos\vartheta
d\varphi_\gamma}.
\ea
It is important to take care of the electron mass $m_e$, e.g. in the
following contribution:
\ba
\label{integral}
\int^{2\pi}_0 d{\varphi_\gamma}\,\frac{1}
{Z_i(R,\cos\vartheta,x,\varphi_\gamma)} 
&=&
\frac{2\pi\sqrt{\lambda_2}}{\sqrt{C_i}},
\ea
\ba
\label{integral2}
C_i &=& s^2 a_i x^2 - 2 s b_i x  + c_i,
\\
a_i &=& s^2 ( z_i^2-R\,\eta_0^2),
\\
b_i &=& s^3[ R\, z_i\,(1- z_i)-\frac{1}{2}\,R(1-R)\,\eta_0^2],
\\
c_i &=& s^4\,R^2\,(1- z_i)^2,
\\
z_{1(2)} &=& \frac{1\mp\beta_0\cos\vartheta}{2}
+ R\frac{1\pm\beta_0\cos\vartheta}{2}
\\
\eta_0^2 &=& 1-\beta_0^2;\quad \beta_0^2 = 1-\frac{4 m_e^2}{s}.
\ea
Be the next integration that over $x$ with limits (\ref{ibound}).
Consider e.g. the following integral:
\ba 
\label{integral3}
I_i^0(R,\cos\vartheta)
&=&
\int\limits_{x_{min}}^{x_{max}}
\frac{sd x}{\sqrt{C_i}} 
\nl
&=&
\frac{1}{\sqrt{a_i}}
\left.
\ln\left[\sqrt{a_i} C_i^{\frac{1}{2}}+ (s a_i x - b_i) \right]
\right|^{x_{max}}_{x_{min}},
\ea
with
\ba
\label{integral4}
C_i |_{x_{max,min}}
&=& 
\frac{1}{4}\,s^4(1-R)^2
\left[(y_i\pm A z_i)^2+R\,(1-A^2)\,\eta_0^2\right],
\ea

\ba
\label{yi}
y_{1(2)} 
&=& \frac{1\mp\beta_0\cos\vartheta}{2}
- R \frac{1\pm\beta_0\cos\vartheta}{2},
\ea
and
\ba
(s a_i x - b_i)\Biggr|^{x_{max}}_{x_{min}}
&=& (1-R) (y_i \pm A z_i) + O(\eta_0^2).
\ea
We are interested in the limit of vanishing electron mass $m_e$ for the
subsequent integrations.
In this limit there will occur
zeros in arguments of logarithms like that in (\ref{integral3}) at
four locations defined by the conditions:
\ba
\label{yaz}
(y_i \pm A z_i) = 0, ~~~~~i=1,2.
\ea
These zeros appear as functions of $\cos\vartheta$ with parameters
$R$ and $A=A(R)$ at certain values $\cos\vartheta = c_i^{\pm}$ ($i=1,2$):
\ba
\label{c2+}
c_2^{+} &=& \frac{1-R+A(R) (1+R)}{1+R+A(R)(1-R)} \ge 0,
\\
\label{c2-}
c_2^{-} &=& \frac{1-R-A(R) (1+R)}{1+R-A(R) (1-R)},
\\
\label{c1+}
c_1^{-}  &=& - c_2^{-},
\\ 
\label{c1-}
c_1^{+} &=& - c_2^{+}\le 0.
\ea   
The relations $c_2^{+}  \ge c_2^{-}$ and $c_1^{+} \le c_1^{-}$ are
also fulfilled.
In the course of the 
integrations different analytical expressions have to be used in different
kinematical regions when neglecting $m_e$ wherever
possible\footnote{In logarithmic contributions of the type  
$L_e = \ln({s}/{m_e^2})$ and $\ln(1\pm\beta_0\cos\vartheta)$
from collinear photon emission, the electron mass 
has to be taken into account.}.
As a result of all that 
we have to cut the remaining phase space for 
$\cos\vartheta$ (at fixed $R$ and for given $i$) into three different
regions.
It is region I (with the $s'$-cut) where the conditions
(\ref{c2+}) to (\ref{c1-}) become trivial:
\ba 
\label{cI}
c_2^{+}(I) = - c_2^{-}(I) = c_1^{-}(I) = - c_1^{+}(I) =1.
\ea
There, only one case has to be considered 
($-1\leq\cos\vartheta \leq 1$).
In regions II and III,
the differential cross-section  ${d{\sigma}}/({dRd{\cos\vartheta}})$
is a double-sum  over $i=1,2$ and  
has different analytical expressions 
for each combination of the kinematical ranges defined by (\ref{c2+})
to (\ref{c1-}).

The final result for e.g. $I^0_i$ after integration over $x$ setting
$m_e=0$ becomes ($i=1,2$):
\\
(i) for $|\cos\vartheta| < |c^{-}_{i}|$ with $ y_i\pm A z_i>0 $ (case
''$i++$''):  
\ba
\label{I0i}
I_i^0 = \frac{1}{s z_i}\ln\left(\frac{y_i+A z_i}
{y_i-A z_i}\right)
\ea
(ii) for $|c^{-}_{i}|<|\cos\vartheta|<|c^{+}_{i}|$ with $y_i+A z_i> 0$ and
$y_i-A z_i< 0 $ (case ''$i+-$''):
\ba
\label{I0ii}
I_i^0 = \frac{1}{s z_i}\left\{\ln\left[\frac{ z_i^2(y_i+A z_i)
(A z_i-y_i)}{R^2(1-\beta_0^2\cos^2\vartheta)}\right]
+\ln\left(\frac{s}{m_e^2}\right)\right\},
\ea
(iii) for $|\cos\vartheta|>|c^{+}_{i}|$ with $y_i\pm A z_i< 0$ (case
''$i--$''): 
\ba
\label{I0iii}
I_i^0 =  -\frac{1}{s z_i}\ln\left(\frac{y_i+A z_i}
{y_i-A z_i}\right).
\ea
One may show that the resulting number of cases for the angular
distribution, depending on
the value of $\cos\vartheta$ with respect to $c^{\pm}_{i}$ 
and on $R$, is at most four in regions II and III.
These are for $\cos\vartheta\ge 0$ with the abbreviations given 
in eq.~(\ref{I0i}) to~(\ref{I0iii}):
\\
\begin{tabular}{lcc}
\label{case1}
a. && ''$1++$'' combined with ''$2++$'';
\newline\\
b. && ''$1+-$'' combined with ''$2+-$'';
\newline\\
c. && ''$1++$'' combined with ''$2+-$'';
\newline\\
d. && ''$1--$'' combined with ''$2++$''.
\end{tabular}
\\
For $\cos\vartheta<0$ cases c. and d. are exchanged by:
\\
\begin{tabular}{lcc}
\label{case2}
a. && ''$1++$'' combined with ''$2++$'';
\newline\\
b. && ''$1+-$'' combined with ''$2+-$'';
\newline\\
c.' && ''$1+-$'' combined with ''$2++$'';
\newline\\
d.' && ''$1++$'' combined with ''$2--$''.
\end{tabular}
\\
For region I (the $s'$-cut) only case b. is possible. 
This simplification follows from $A=1$; see above.

When integrating over $\cos\vartheta$ within acceptance cut 
boundaries $\pm c$, the distinction
of different regions in phase space has to be repeated where
the cut-off $c$ now plays the role of $\cos\vartheta$.
Depending on the relative position of $c$ with respect to
the values $c_i^{\pm}$ we have to integrate over different 
expressions of the angular distribution (indicated above by 
the distinction of cases a.~to d., or a.~to d.' respectively)
in the remaining $\cos\vartheta$-$R$ phase space.
One finally gets at most four different analytical 
expressions for $\sigma_T^{hard}(R;c,A)$ and six for 
$\sigma_{FB}^{hard}(R;c,A)$ in different regions of phase space.
This is because of symmetric cancellations when 
integrating over $\cos\vartheta$: 
$\sigma_T^{hard} = \int^c_{-c}{d\cos\vartheta}\, 
{d\sigma^{hard}} / {d\cos\vartheta}$,
while
$\sigma_{FB}^{hard} = \left(\int^c_0-\int^0_{-c}\right) 
{d\cos\vartheta}\, 
{d\sigma^{hard}} / {d\cos\vartheta}$.
It is the  additional occurance of $c=0$ in the definition of
$\sigma_{FB}^{hard}$ that leads to more cases.

If no acceptance cut is applied, $c=1$, only one case remains for 
$\sigma_T^{hard} = \sigma(1) - \sigma(-1)$ 
in regions II and III 
-- because then $-1< c^{\pm}_{i} < 1$ (see cases d. and d'~above) -- while 
for $\sigma_{FB}^{hard}$ two of them are left
because the additional integrated contributions from 
$\cos\vartheta=0$ depend on whether 
$c^{-}_{2}>0$ or not ($c^{-}_{2}=-c^{-}_{1}$):
$\sigma_{FB}^{hard} = \sigma(1)-2 \sigma(0)+ \sigma(-1)$.
The conditions (\ref{c2-}) and   (\ref{c1+}) are fulfilled for
$\cos\vartheta = 0$ with 
\ba
\label{a0}
A_0 &=& \frac{1-R}{1+R},
\ea
so that, depending on $A(R)$, one or the other analytical expression has
to be used.
This was practised in \cite{Christova:1999cc}. 

One can check that 
the integrated results $\sigma_{T,FB}^{hard}$ are continuous when 
$c\rightarrow c_i^{\pm}$ while 
$d\sigma^{hard}/d\cos\vartheta$ can be regularised at
$\cos\vartheta=c_i^{\pm}$ 
taking the exact logarithmic results in $m_e$ for the integrals.  

One can also reassure oneself that the contributions proportional to
the Born cross-section $\sigma^0$ and Born asymmetry $A_{FB}$ are 
(anti)symmetric respectively as it should be for the one loop
corrected initial-state results. 
We will come to this in the next section.

\bigskip

The phase-space splitting discussed above has also an influence 
on the initial-final state interference corrections since there the 
initial-state propagators with $Z_{1,2}^{-1}$ appear linearly.
They will not contribute in the final-state contributions so that the 
phase-space splitting is not necessary there.

\subsection{\large \bf Initial-state radiation
\label{secini}
}
We systematically compared numerical predictions of \zf\ v.5.20 and 
\zf\ v.6.11 with default flag settings.
Version 5.20 was used as released, while version 6.11 was prepared
such that the changes due to the recalculation of initial-state
corrections, final-state 
corrections, their interferences, and the net effect could be isolated.
We begin with a study of the changes related to {\it initial-state
radiation}. 

For $\sigma_T$, the changes 
are at most one unit in the fifth digit at LEP~1 energies
and thus considered to be completely negligible.
In Table \ref{tabini10} 
we show the corresponding shifts of predictions for 
$A_{FB}$ for two acollinearity cuts,
$\theta_{\rm acol}<10^{\circ},25^{\circ}$, 
and
three different acceptance cuts,
$\vartheta_{acc}=0^{\circ},20^{\circ},40^{\circ}$.
The changes are also less than the theoretical accuracies demanded.
We checked that our numbers for  {\tt ICUT} = 0 agree with 
the \zf\ predictions shown in Tables 26 and 27 of \cite{Bardin:1999gt}.

\begin{figure}[t] 
\begin{flushleft}
  \begin{tabular}{ll}
  \mbox{%
  \epsfig{file=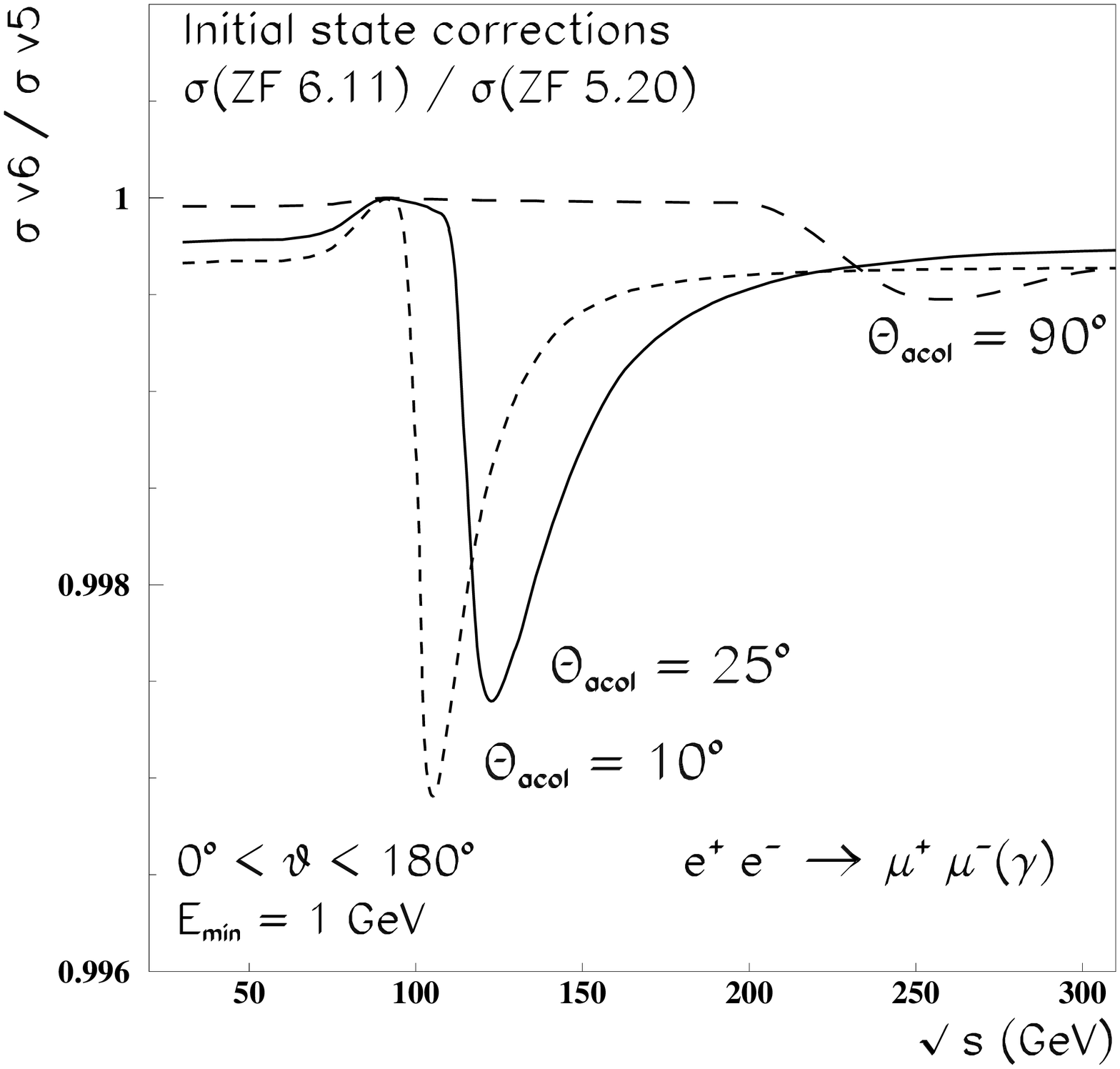,width=7.7cm}}%
&
  \mbox{%
  \epsfig{file=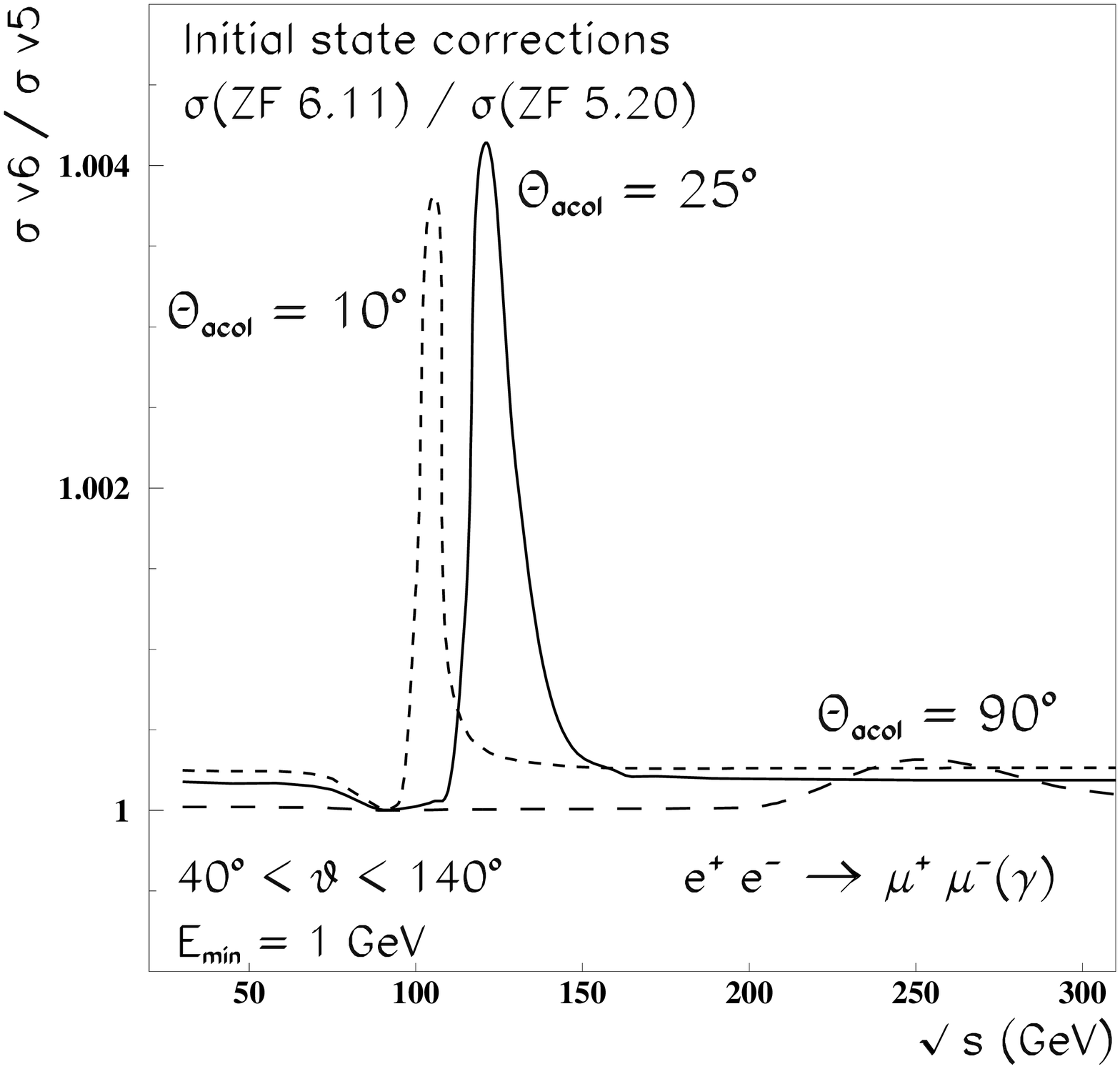,width=7.7cm}}%
\end{tabular}
\caption
{\sf
Ratios of muon-pair production cross-sections predicted from \zf\ 
v.6.11 and v.5.20 without and with acceptance cut and with three different
acollinearity cuts: $\theta_{\rm acol} < 10^{\circ}, 25^{\circ}, 90^{\circ}$;
$E_{min}=1$ GeV;  programs differ by initial-state radiation.
\label{fig26sigc0} 
}
\end{flushleft}
\end{figure}

\begin{figure}[t] 
\begin{flushleft}
  \begin{tabular}{ll}
  \mbox{%
  \epsfig{file=%
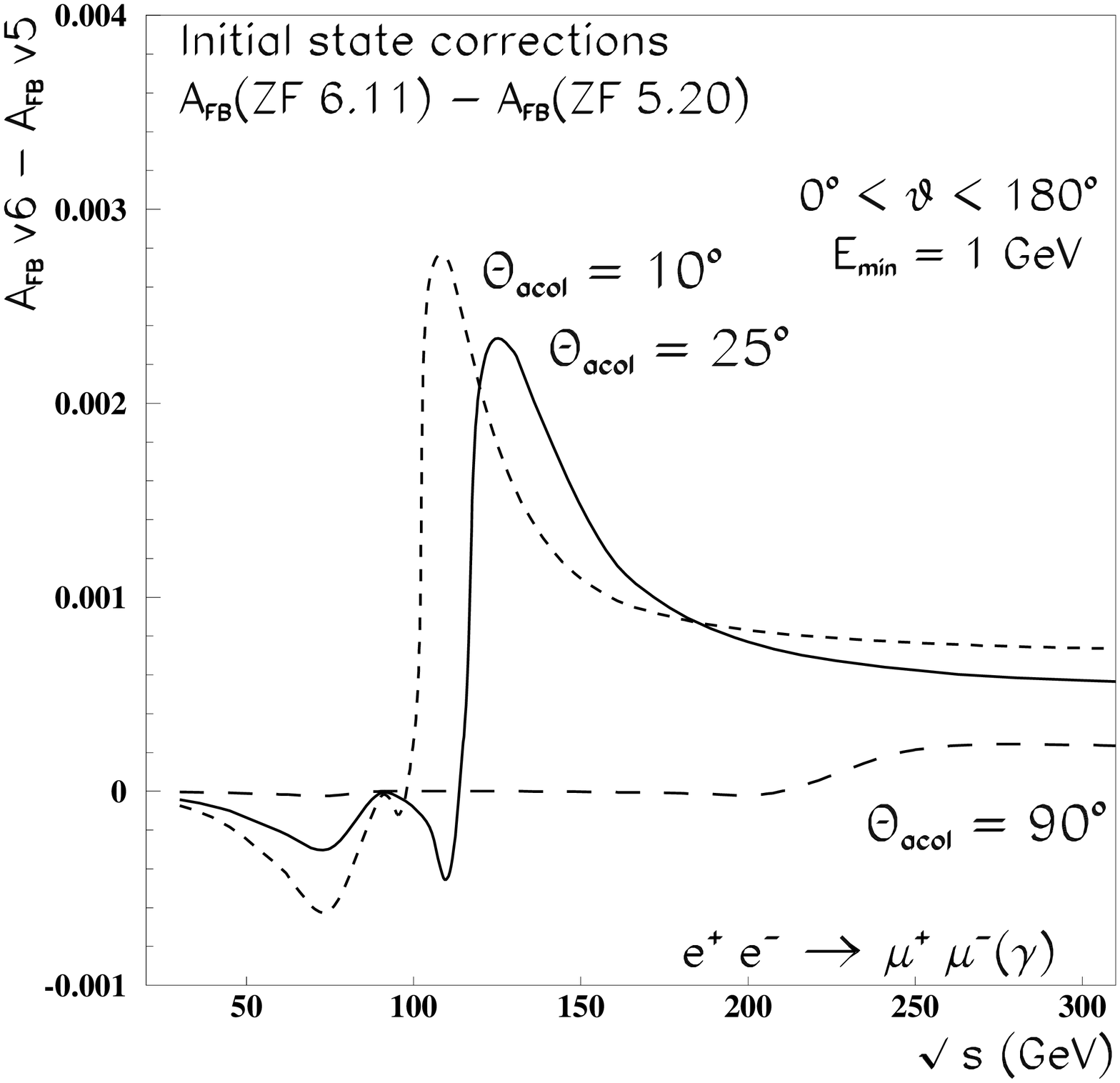
           ,width=7.7cm   
         }}%
&
  \mbox{%
\epsfig{file=%
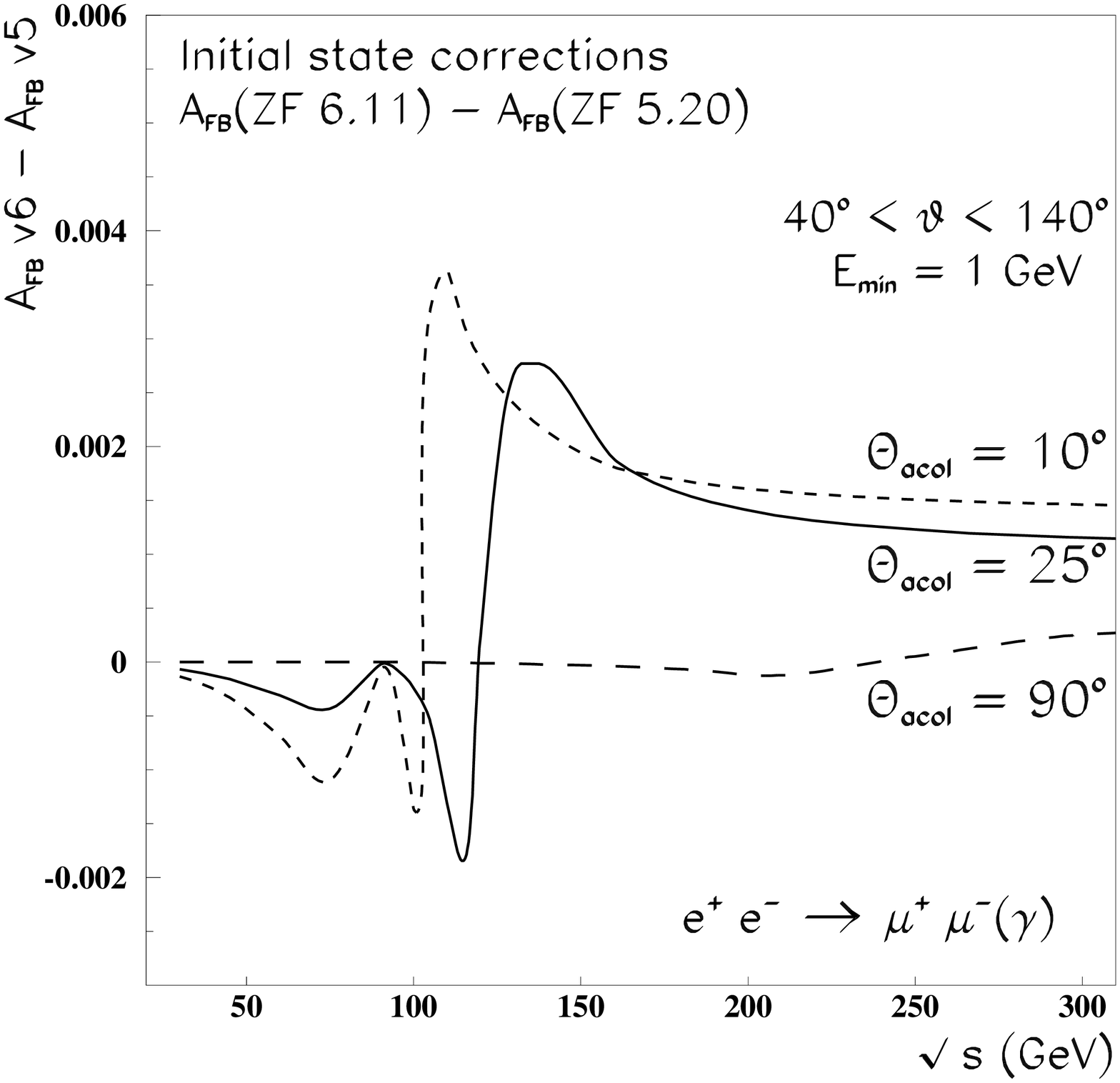
           ,width=7.7cm   
         }}%
\end{tabular}
\caption
{\sf 
Differences of muonic forward-backward asymmetries
predicted from \zf\  
v.6.11 and v.5.20 without and with acceptance cut and with three different
acollinearity cuts: $\theta_{\rm acol} < 10^{\circ}, 25^{\circ}, 90^{\circ}$;
$E_{min}=1$ GeV; programs differ by initial-state radiation.
\label{fig26afbc0} 
}
\end{flushleft}
\end{figure}

In Figure \ref{fig26sigc0}
the ratio of $\sigma_T$ from \zf\ v.6.11 and v.5.20
and in Figure \ref{fig26afbc0}
the difference of  $A_{FB}$ are shown in a wider energy range.
While at energies slightly above the $Z$ peak
the differences of the predictions show local peaks, at
LEP~2 energies and beyond they are negligible for  $\sigma_T$ and  
amount to only 0.1\% 
 -- 0.2\% 
for $A_{FB}$.
The peaking structures disappear at energies for which the radiative
return is prohibited by the cuts.
In dependence on the acollinearity
cut, this happens for energies $s > s^{min}$. 
The $s^{min}$ is calculated in Appendix \ref{app-ret}. 

\begin{table}[ht]
\begin{center}
\renewcommand{\arraystretch}{1.1}
 \begin{tabular}{|c||c|c|c|c|c|}
\hline
\multicolumn{6}{|c|}{$\afba{\flm}$ with $\theta_{\rm acol}<10^{\circ}$} \\
\hline
 $\theta_{\rm acc}$& $\mz - 3$ & $\mz - 1.8$ & $\mz$ & $\mz + 1.8$ &
$\mz + 3$  \\ 
\hline \hline
$0^{\circ}$
      &-0.28462  &-0.16916  & 0.00024  & 0.11482  & 0.16063 \\
      &-0.28453  &-0.16911  & 0.00025  & 0.11486  & 0.16071 \\
\hline
$20^{\circ}$
       &-0.27521  &-0.16355  & 0.00032  & 0.11141  & 0.15602  \\
       &-0.27506  &-0.16347  & 0.00035  & 0.11148  & 0.15616  \\
\hline
$40^{\circ}$
       &-0.24230  &-0.14398  & 0.00045  & 0.09881  & 0.13868  \\
       &-0.24207  &-0.14386  & 0.00050  & 0.09893  & 0.13891  \\
\hline 
\hline 
\multicolumn{6}{|c|}{$\afba{\flm}$ with $\theta_{\rm acol}<25^{\circ}$} \\
\hline
 $\theta_{\rm acc}$
& $\mz - 3$ & $\mz - 1.8$ & $\mz$ & $\mz + 1.8$ & $\mz + 3$  \\
\hline \hline
  $0^{\circ}$
        &-0.28651  &-0.17051  &-0.00043  & 0.11292  & 0.15680  \\
        &-0.28647  &-0.17049  &-0.00043  & 0.11293  & 0.15682  \\
\hline
$20^{\circ}$
        &-0.27727  &-0.16499  &-0.00038  & 0.10942  & 0.15201  \\
        &-0.27722  &-0.16497  &-0.00037  & 0.10944  & 0.15204  \\
\hline
$40^{\circ}$
        &-0.24452  &-0.14549  &-0.00027  & 0.09675  & 0.13449  \\
        &-0.24445  &-0.14545  &-0.00026  & 0.09678  & 0.13454  \\
\hline 
\end{tabular}
\caption{\sf
Comparison of \zf\ v.6.11 (first row) with \zf\ v.5.20 (second row)
for the muonic forward-backward asymmetry with
  angular acceptance cut ($\theta_{\rm acc}=0^{\circ},20^{\circ},40^{\circ}$)
  and acollinearity cut ($\theta_{\rm acol}<10^{\circ},25^{\circ}$);
$\mz=91.1867$ (GeV).
The initial-final state interference is switched off and only
initial-state radiation is corrected.
\label{tabini10}
}
\end{center}
\end{table}

{For} the case of initial-state radiation, we were able to trace back the
reason of the numerical inaccuracies related to the acollinearity cut
of \zf\ below version 6. 
It is the result of leaving out a certain class of non-logarithmic, simple
terms  of order  $O(\alpha)$.
{For} $\sigma_T$, polynomials proportional to $\cos\vartheta$ (and their
integrals) are concerned, and for $\sigma_{FB}$ polynomials of the type
$(a+b\cos^2\vartheta)$ (and their integrals).
At first glance the corresponding contributions seem to vanish for
symmetric acceptance cuts. 
But this is not the case!
As we explained in Section \ref{sec:cases}, the cross-section formulae 
lose the usual simple symmetry/anti-symmetry behaviour under the
transformation $\cos\vartheta \leftrightarrow (-\cos\vartheta)$ in
regions II and III of the phase space since different analytical
expressions may be needed depending on the location of the parameters
$c_i^{\pm}$ describing the solutions of (\ref{yaz}). 
Then, the symmetry behaviour as a function of
$\cos\vartheta$ is ''hidden'' since different regions contribute differently to
the net result.

\subsection{\large \bf Initial-final state interference
\label{interference}
}
The \zf\ v.5 predictions of photonic corrections from the {\it
initial-final state interference} also receive modifications due to the
recalculation for the versions v.6.  
The explanation given in Section \ref{secini} for the case of
initial-state radiation is also applicable for a part of the deviations
here. The codings for the initial-final state interference also show
additional deviations in the hard photonic corrections and the
resulting numerical differences are much larger.   

\begin{sidewaystable}
\renewcommand{\arraystretch}{1.1}
\begin{tabular}{|c||c|c|c|c|c||c||c|c|c|c|c|}
\hline
\multicolumn{6}{|c|}
{$\sigma_{\flm}\,$[nb] with $\theta_{\rm acol}<10^{\circ}$} &
\multicolumn{6}{|c|}
{$\sigma_{\flm}\,$[nb] with $\theta_{\rm acol}<25^{\circ}$} \\
\hline
$\theta_{\rm acc}$
& $\mz - 3$ & $\mz - 1.8$ & $\mz$ & $\mz + 1.8$ & $\mz + 3$  &
$\theta_{\rm acc}$
& $\mz - 3$ & $\mz - 1.8$ & $\mz$ & $\mz + 1.8$ & $\mz + 3$  \\
\hline\hline
{\tt Z6} \hfill $0^{\circ}$
  & 0.21928  & 0.46285  & 1.44780  & 0.67721  & 0.39360 &
{\tt Z6} \hfill $0^{\circ}$ 
  & 0.22328  & 0.46968  & 1.46598  & 0.68688  & 0.40031 \\
  & 0.21772  & 0.46082  & 1.44776  & 0.67898  & 0.39489 &
  & 0.22228  & 0.46836  & 1.46602  & 0.68816  & 0.40128 \\
  &-7.16     &-4.41     &-0.03     &+2.60     &+3.27    &
  &-4.51     & -2.82    & +0.03    & +1.86    & +2.41   \\
\hline
{\tt Z5} \hfill $0^{\circ}$  
  & 0.21928  & 0.46285  & 1.44781  & 0.67722  & 0.39361 &
{\tt Z5} \hfill $0^{\circ}$ 
  & 0.22328  & 0.46968  & 1.46598  & 0.68688  & 0.40031 \\
  & 0.21852  & 0.46186  & 1.44782  & 0.67814  & 0.39429 &
  & 0.22281  & 0.46905  & 1.46603  & 0.68754  & 0.40081 \\
  &-3.48     &-2.14     &+0.01     &+1.36     &+1.72    &
  &-2.11     &-1.34     &+0.03     &+0.96     &+1.25    \\
\hline\hline
{\tt Z6} \hfill $20^{\circ}$
  & 0.19987  & 0.42205  & 1.32053  & 0.61756  & 0.35881 &
{\tt Z6} \hfill $20^{\circ}$ 
  & 0.20357  & 0.42834  & 1.33718  & 0.62647  & 0.36505 \\  
  & 0.19869  & 0.42046  & 1.32018  & 0.61877  & 0.35972 &
  & 0.20281  & 0.42729  & 1.33689  & 0.62731  & 0.36572 \\
  & -5.96    &-3.79     &-0.27     &+1.95     &+2.53    &
  & -3.74    & -2.46    & -0.21    & +1.35    & +1.83   \\
\hline
{\tt Z5} \hfill $20^{\circ}$ 
  & 0.19987  & 0.42205  & 1.32053  & 0.61756  & 0.35881 &
{\tt Z5} \hfill $20^{\circ}$ 
  & 0.20357  & 0.42833  & 1.33718  & 0.62647  & 0.36505 \\
  & 0.19892  & 0.42075  & 1.32021  & 0.61857  & 0.35959 &
  & 0.20321  & 0.42781  & 1.33689  & 0.62684  & 0.36536 \\
  &-4.78     &-3.09     &-0.24     &+1.63     &+2.17    &
  &-1.77     &-1.22     &-0.22     &+0.59     &+0.85    \\
\hline\hline
{\tt Z6} \hfill $40^{\circ}$ 
  & 0.15032  & 0.31760  & 0.99416  & 0.46475  & 0.26983 &
{\tt Z6} \hfill $40^{\circ}$ 
  & 0.15318  & 0.32243  & 1.00682  & 0.47164  & 0.27477 \\ 
  & 0.14974  & 0.31675  & 0.99349  & 0.46515  & 0.27019 &
  & 0.15280  & 0.32183  & 1.00619  & 0.47188  & 0.27502 \\
  &-3.88     &-2.72     &-0.67     &+0.87     &+1.32    &
  & -2.48    & -1.88    & -0.62    & +0.51    & +0.91   \\ 
\hline
{\tt Z5} \hfill $40^{\circ}$ 
  & 0.15032  & 0.31760  & 0.99415  & 0.46474  & 0.26983 &
{\tt Z5} \hfill $40^{\circ}$ 
  & 0.15318  & 0.32243  & 1.00682  & 0.47164  & 0.27477 \\
  & 0.14978  & 0.31680  & 0.99350  & 0.46511  & 0.27016 &
  & 0.15287  & 0.32192  & 1.00619  & 0.47180  & 0.27496 \\
  &-3.61     &-2.53     &-0.65     &+0.80     &+1.22    &
  &-2.03     &-1.58     &-0.63     &+0.34     &+0.69    \\
\hline 
\end{tabular}
\caption
{\sf
Comparison of \zf\ v.6.11 (first row) with \zf\ v.5.20 (second row)
for  muon-pair production cross-sections
with angular acceptance cuts 
($\theta_{\rm acc}=0^{\circ},20^{\circ},40^{\circ}$) and acollinearity
cut ($\theta_{\rm acol}<10^{\circ},25^{\circ}$).
First row is without initial-final state
interference, second row with,
third row the relative effect of that interference in per mil.
Final-state treated as in v.5.20.
\label{tab10sigifi}
}
\end{sidewaystable}

In Table \ref{tab10sigifi}
we show the shifts of predictions for
muon-pair production due to the initial-final state interference for
two choices of the acollinearity cut.
Table \ref{tab10afbifi}
shows the corresponding effects for the
forward-backward asymmetry.
The two tables are the analogues to Tables 37--40 of
\cite{Bardin:1999gt}, where 
{\tt TOPAZ0} v.4.3 and \zf\ v.5.20 were compared. 
At the $Z$ peak, the predictions for the influence of the
initial-final state 
interference from \zf\ v.5.20 and \zf\ v.6.11 deviate from each other only
negligibly, with maximal deviations of up to 0.015\%.
At the wings, the situation is quite different; we observe
deviations of up to several per mil for cross-sections and 
up to a per mil for asymmetries.
The deviations between the two codings decrease if the
acollinearity cut is weakened.

\begin{sidewaystable}
\renewcommand{\arraystretch}{1.1}
\begin{tabular}{|c||c|c|c|c|c||c||c|c|c|c|c|}
\hline
\multicolumn{6}{|c||}{$\afba{\flm}$ with $\theta_{\rm acol}<10^{\circ}$} &
\multicolumn{6}{|c|}{$\afba{\flm}$ with $\theta_{\rm acol}<25^{\circ}$} \\
\hline
$\theta_{\rm acc}$
& $\mz - 3$ & $\mz - 1.8$ & $\mz$ & $\mz + 1.8$ & $\mz + 3$  &
$\theta_{\rm acc}$
& $\mz - 3$ & $\mz - 1.8$ & $\mz$ & $\mz + 1.8$ & $\mz + 3$  \\
\hline\hline
{\tt Z6} \hfill $0^{\circ}$  
  & -0.28462 & -0.16916  & 0.00024  & 0.11482  & 0.16063 &
{\tt Z6} \hfill $0^{\circ}$ 
  & -0.28651 & -0.17051 & -0.00043  & 0.11292  & 0.15680 \\ 
  & -0.28187 & -0.16689  & 0.00083  & 0.11379  & 0.15907 &
  & -0.28554 & -0.16960 & -0.00000  & 0.11285  & 0.15669 \\
  & +2.75    & +2.27     & +0.60    &-1.03     &-1.56    &
  & +0.97    & +0.91    & +0.43     & -0.06    & -0.11   \\
\hline
{\tt Z5} \hfill $0^{\circ}$  
  & -0.28453 & -0.16911  & 0.00025  & 0.11486  & 0.16071 &
{\tt Z5} \hfill $0^{\circ}$ 
  & -0.28647 & -0.17049 & -0.00043  & 0.11293  & 0.15682 \\
  & -0.28282 & -0.16783  & 0.00070  & 0.11475  & 0.16059 &
  & -0.28555 & -0.16975 & -0.00005  & 0.11307  & 0.15701 \\
  & +1.71    & +1.28     &+0.45     &-0.11     &-0.12    &
  & +0.92    & +0.74    & +0.48     &+0.14     &+0.19    \\
\hline \hline
{\tt Z6} \hfill $20^{\circ}$
  & -0.27521 & -0.16355  & 0.00032  & 0.11141  & 0.15602 &
{\tt Z6} \hfill $20^{\circ}$ 
  & -0.27727 & -0.16499 & -0.00038  & 0.10942  & 0.15201 \\ 
  & -0.27285 & -0.16167  & 0.00080  & 0.11053  & 0.15467 &
  & -0.27659 & -0.16436 & -0.00006  & 0.10943  & 0.15199 \\
  & +2.35    & +1.88     &+0.47     & -0.89    &-1.35    &
  & +0.68    & +0.63    & +0.32     & +0.00    & -0.02   \\
\hline
{\tt Z5} \hfill $20^{\circ}$  
  & -0.27506 & -0.16347  & 0.00035  & 0.11148  & 0.15616 &
{\tt Z5} \hfill $20^{\circ}$ 
  & -0.27722 & -0.16497 & -0.00037  & 0.10944  & 0.15204 \\
  & -0.27408 & -0.16261  & 0.00070  & 0.11133  & 0.15594 &
  & -0.27657 & -0.16447 & -0.00009  & 0.10963  & 0.15229 \\
  & +0.98    & +0.86     &+0.35     &-0.15     &-0.22    &
  & +0.65    & +0.50    & +0.28     &+0.19     &+0.25    \\
\hline\hline
{\tt Z6} \hfill $40^{\circ}$
  & -0.24230 & -0.14398  & 0.00045  & 0.09881  & 0.13868 &
{\tt Z6} \hfill $40^{\circ}$ 
  & -0.24452 & -0.14549 & -0.00027  & 0.09675  & 0.13449 \\
  & -0.24063 & -0.14277  & 0.00073  & 0.09825  & 0.13780 &
  & -0.24423 & -0.14527 & -0.00010  & 0.09687  & 0.13464 \\ 
  & +1.67    & +1.22     & +0.28    & -0.56    & -0.88   &
  & +0.29    & +0.22    & +0.17     & +0.12    & +0.15   \\
\hline
{\tt Z5} \hfill $40^{\circ}$ 
  & -0.24207 & -0.14386  & 0.00050  & 0.09893  & 0.13891 &
{\tt Z5} \hfill $40^{\circ}$ 
  & -0.24445 & -0.14545 & -0.00026  & 0.09678  & 0.13454 \\
  & -0.24151 & -0.14343  & 0.00069  & 0.09890  & 0.13888 &
  & -0.24444 & -0.14542 & -0.00011  & 0.09700  & 0.13483 \\
  & +0.56    & +0.43     &+0.19     &-0.03     &-0.03    &
  & +0.01    & +0.03    & +0.15     &+0.22     &+0.29    \\
\hline 
\end{tabular}
\caption
{\sf
Comparison of \zf\ v.6.11 (first row) with \zf\ v.5.20 (second row)
for the muonic forward-backward asymmetry 
with angular acceptance cuts ($\theta_{\rm
    acc}=0^{\circ},20^{\circ},40^{\circ}$) and acollinearity cut 
  ($\theta_{\rm acol}<10^{\circ},25^{\circ}$). First row is without
initial-final state 
interference, second row with,
third row the relative effect of that interference in per mil.
Final-state treated as in v.5.20.
\label{tab10afbifi}
}
\end{sidewaystable}

\begin{figure}[t] 
\begin{flushleft}
\begin{tabular}{ll}
  \mbox{%
 \epsfig{file=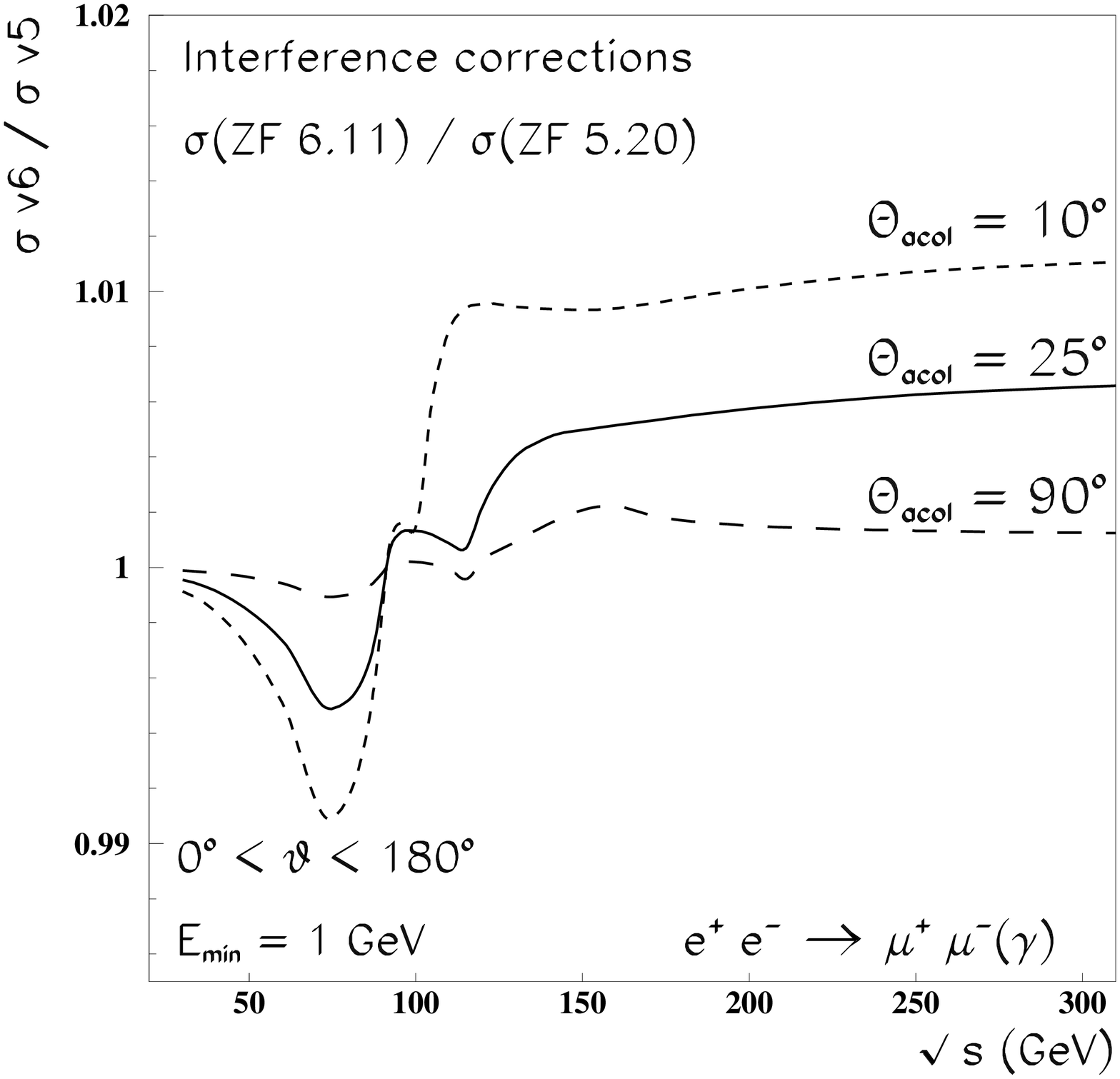
           ,width=7.7cm   
         }}%
&
  \mbox{%
 \epsfig{file=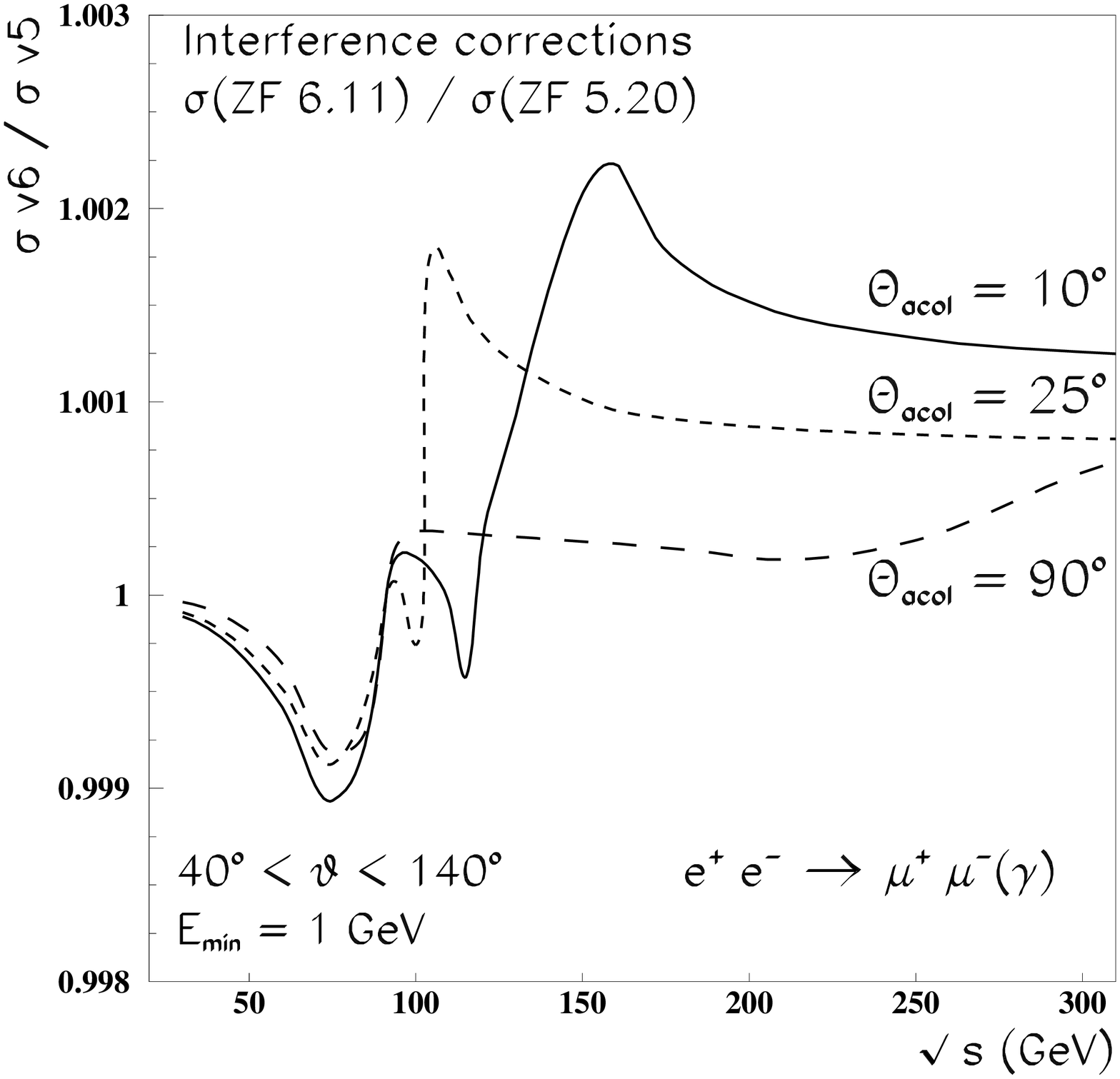
           ,width=7.7cm   
         }}%
\end{tabular}
\caption
{\sf
Ratios of muon-pair production cross-sections predicted from \zf\ 
v.6.11 and v.5.20 without and with acceptance cut and with three different
acollinearity cuts: $\theta_{\rm acol} < 10^{\circ}, 25^{\circ}, 90^{\circ}$;
$E_{min}=1$ GeV; programs differ by initial-final state interference.
\label{fig3739c0} 
}
\end{flushleft}
\end{figure}

\begin{figure}[t] 
\begin{flushleft}
\begin{tabular}{ll}
  \mbox{%
\epsfig{file=%
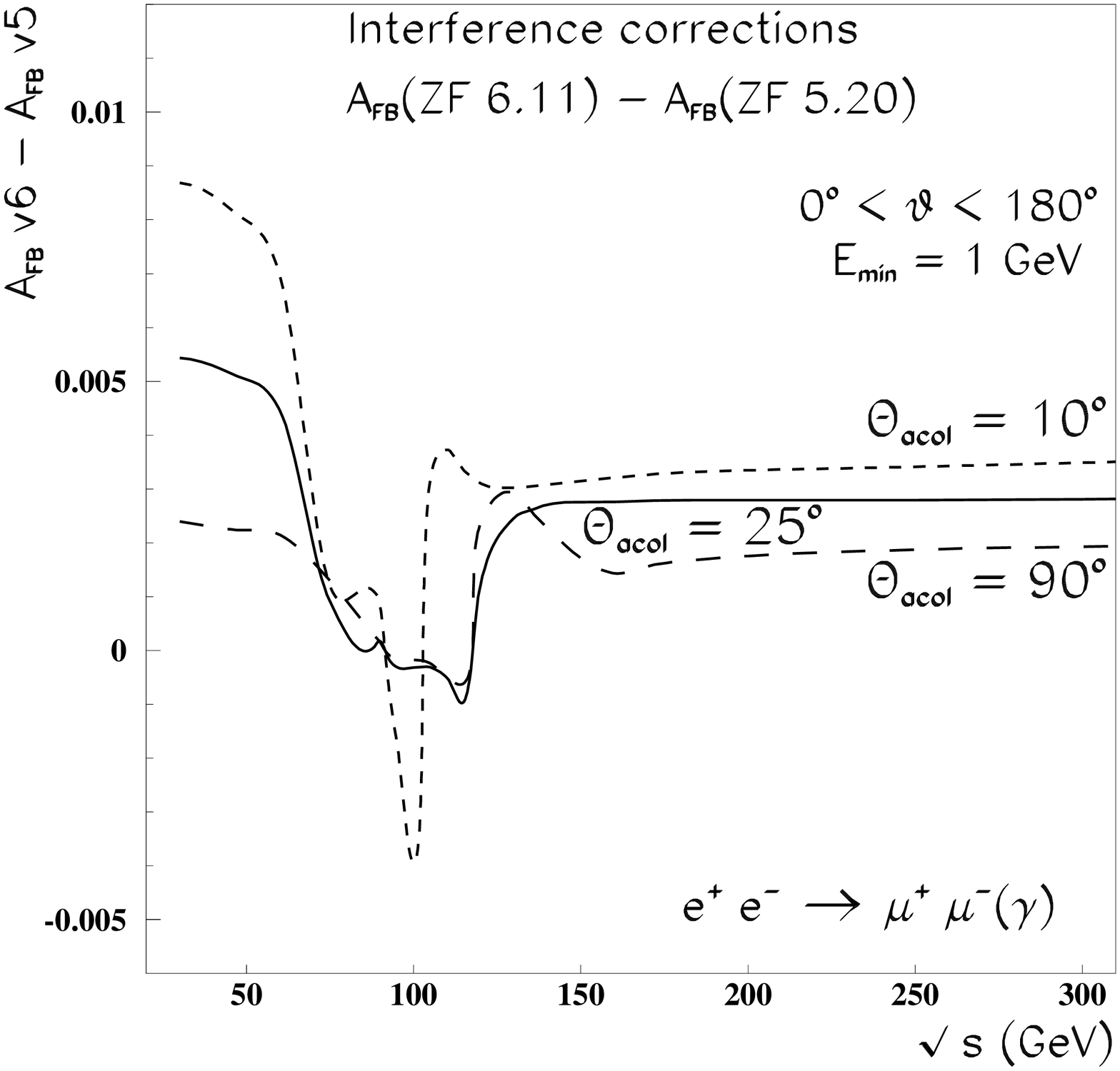
           ,width=7.7cm   
         }}%
&
  \mbox{%
\epsfig{file=%
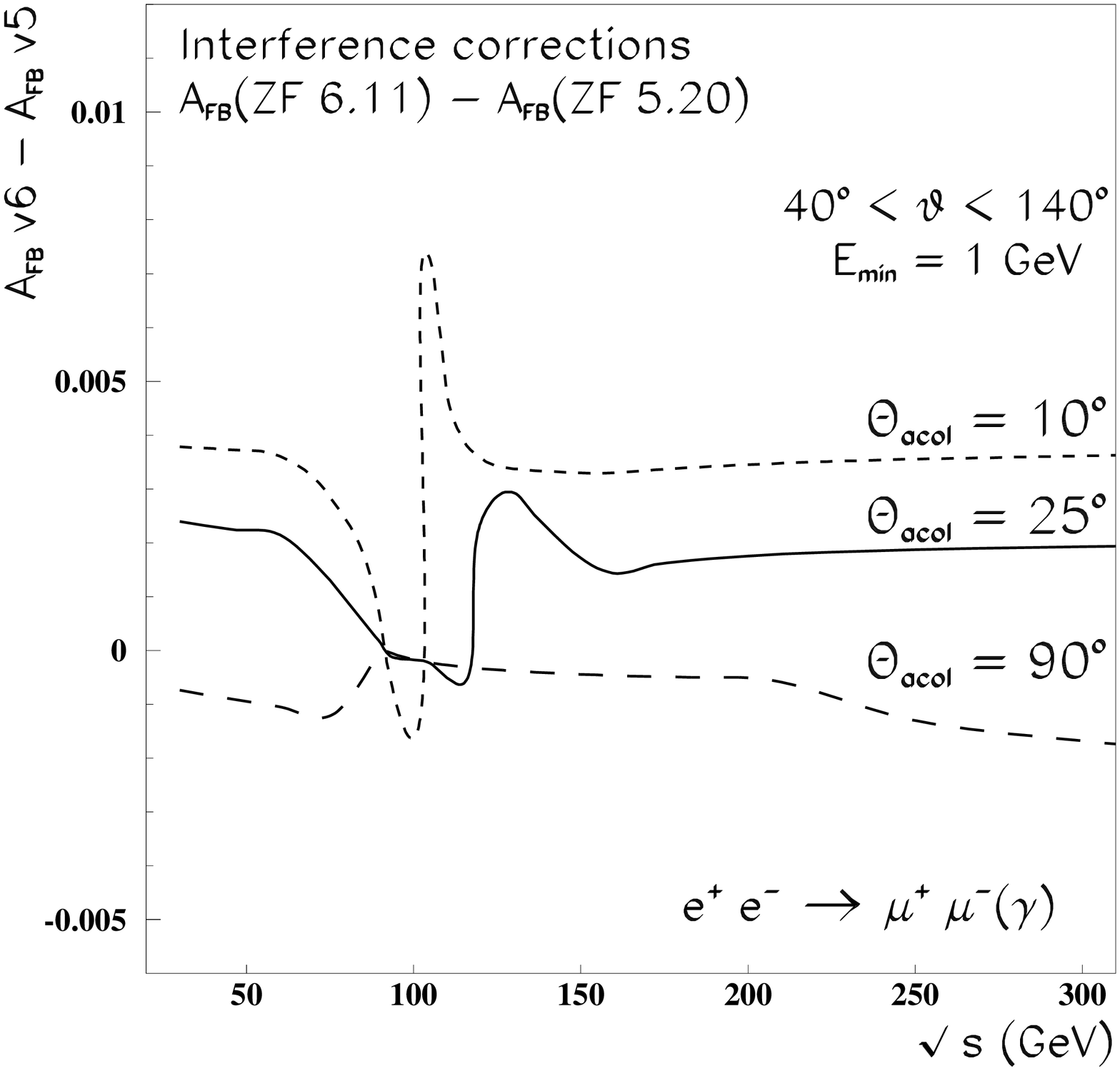
           ,width=7.7cm   
         }}%
\end{tabular}
\caption
{\sf
Differences of muonic forward-backward asymmetries
predicted from \zf\  
v.6.11 and v.5.20 without and with acceptance cut and with three different
acollinearity cuts: $\theta_{\rm acol} < 10^{\circ}, 25^{\circ}, 90^{\circ}$;
$E_{min}=1$ GeV; programs differ by initial-final state interference.
\label{fig3840c0} 
}
\end{flushleft}
\end{figure}

In Figure \ref{fig3739c0}
the corresponding shifts are shown 
for $\sigma_T$ (relative units) for
$\theta_{acc}=0^{\circ},40^{\circ}$,
and
in Figure
\ref{fig3840c0} 
with
$\theta_{acc}=0^{\circ},40^{\circ}$,
for
$A_{FB}$ (absolute units) 
in a wide range of energies.
{For} the cross-sections, the deviations may reach at most up to 
1\%
at LEP~2 energies, while for asymmetries they stay below 0.5\% 
there.
Both shifts are more than the precision we aim at for the theoretical
predictions. 

\subsection{\large \bf Final-state corrections
\label{final}
}
{For} the case of final-state radiation, common soft-photon
exponentiation together with initial-state radiation 
is foreseen in \zf. 
{For} an $s'$-cut, \zf\ follows \cite{Montagna:1993mf}.
As may be seen from \cite{Bilenkii:1989zg} (for the angular
distributions) or from \cite{Christova:1999cc} 
(for integrated observables), the predictions for common
soft-photon exponentiation include one additional integration, namely 
that over the invariant mass of the final-state fermion pair at a
given reduction of $s$ into $s'$ after initial-state
radiation\footnote{With acollinearity cut, there remains some
arbitraryness in the choice of the region with exponentiation.
We did not change what was realised in \zf\ v.5:
Initial-state radiation is exponentiated for $R > \max(R_E,R_{\xi})$
and the final-state radiation, at given $R$, for 
$R' > \max(R_{min}/R,R_E)$. 
A preferred condition might be $R' > \max(R_{min}/R,R_E,R_{\xi})$. 
In this case, for $R_{min}/R < R_{\xi}$, the
non-exponentiated hard photonic corrections from region III would have to be 
left out in order to avoid double counting.
}.
This additional integration is called from subroutine {\tt FUNFIN}
and is treated in \zf\ in a mixed approach.
{For} not too involved integrands, the integration was performed
analytically, while
the integration of two logarithms is being done
numerically using the Lagrange interpolating formula for the integrand 
(subroutine {\tt INTERP} and functions {\tt FAL1} and {\tt FAL2}).
A lattice of 20 points is used with subroutine {\tt INTERP}, and
defining a more dense lattice gave no improvements.
In \zf\ v.5, we found a wrong sign of the term $D_2$ in variable
{\tt SFIN} in subroutine  
{\tt FUNFIN}  and a wrong definition of the argument of function {\tt FAL2}.
The former contributes to   $\sigma_T$, the latter to  $\sigma_{FB}$.
Additionally, we checked the numerical stability related to the numerical
integration.
We see also no problem related to a neglect of 
some  $\cos\vartheta$ dependent terms for  $\sigma_T$ in subroutine
{\tt FUNFIN}. 

The variables mentioned above are defined in \cite{DESY99070}. 

\bigskip

{For} LEP~1, the numerical outcome of our minor improvements is shown in 
Table \ref{tab10fin} (for $\theta_{\rm acol}<10^{\circ},25^{\circ}$) for
$A_{FB}$ at several different acceptance cuts:
$\vartheta_{acc}=0^{\circ},20^{\circ},40^{\circ}$.
Again, our numbers for  {\tt ICUT} = 0 agree with those shown in
Tables 26 and 27 of \cite{Bardin:1999gt}.
All the changes are though visible, but negligible.
{For} the cross-sections, the differences are completely negligible and
not tabulated here.

\begin{figure}[th] 
\begin{flushleft}
\begin{tabular}{ll}
  \mbox{%
  \epsfig{file=%
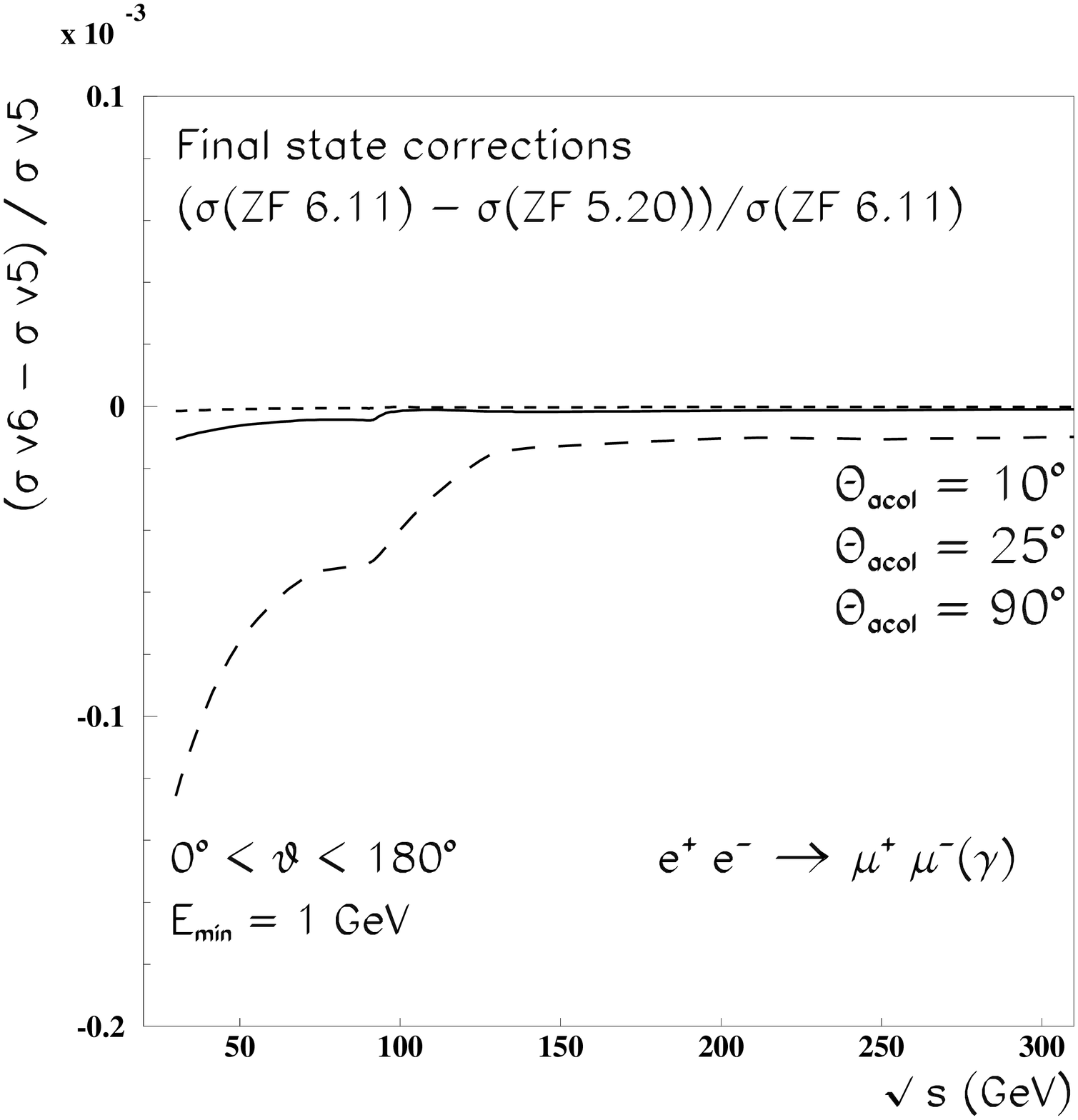
           ,width=7.7cm   
         }}%
&
  \mbox{%
  \epsfig{file=%
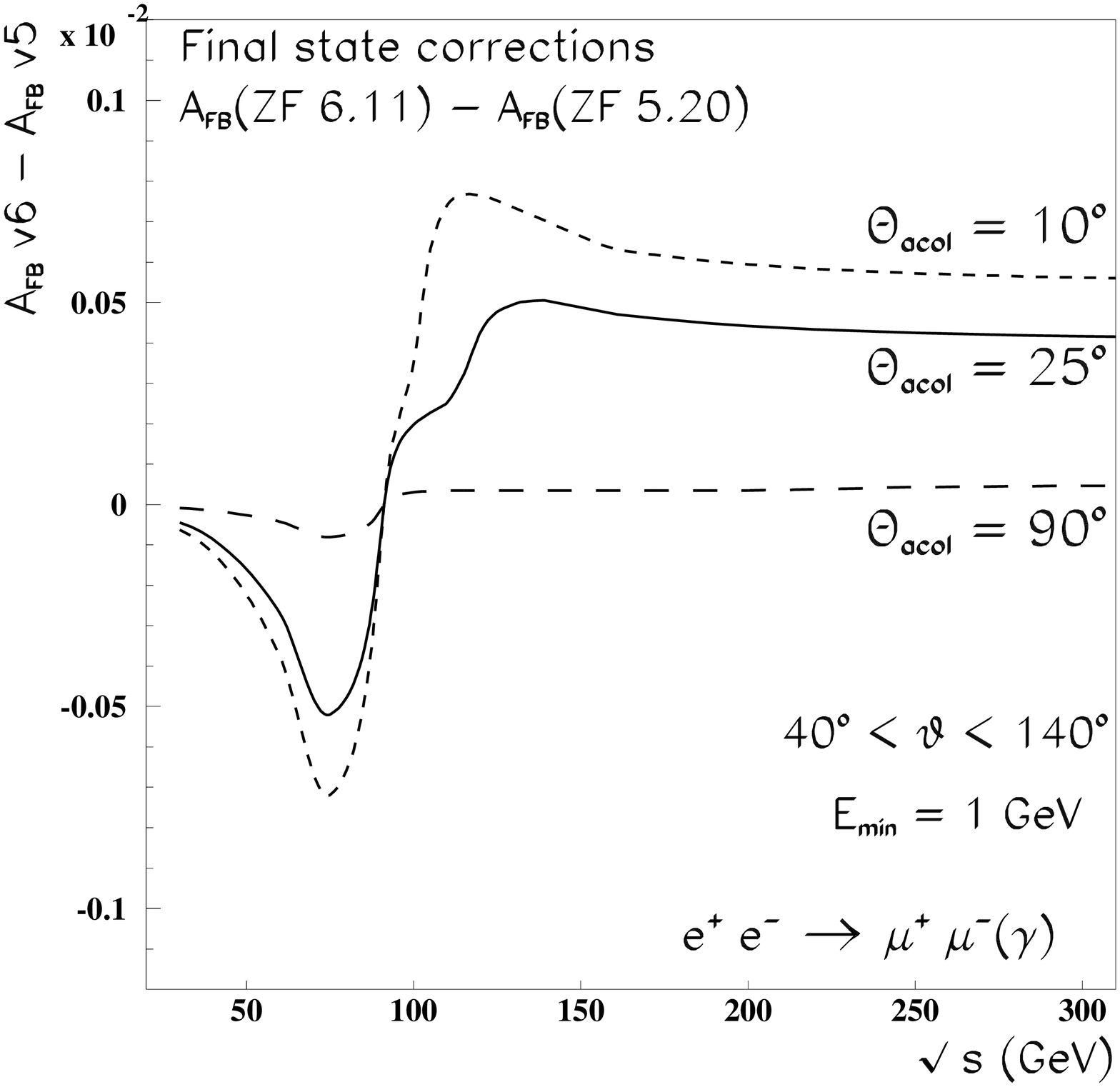
           ,width=7.7cm   
         }}%
\end{tabular}
\caption
{\sf
Ratios of muon-pair production cross-sections and differences of 
forward-backward asymmetries predicted from \zf\ 
v.6.11 and v.5.20 without and with acceptance cut and with three different
acollinearity cuts: $\theta_{\rm acol} < 10^{\circ}, 25^{\circ}, 90^{\circ}$,
$E_{min}=1$ GeV;  
programs differ by final-state radiation.
\label{fig26sigc0F} 
}
\end{flushleft}
\end{figure}

In Figure \ref{fig26sigc0F}, 
the corresponding shifts from \zf\ v.5.20 to v.6.11 are shown 
for $\sigma_T$ (relative units) and 
for $A_{FB}$ (absolute units)
in a wide energy range for the case without acceptance cut.
We see that the deviations are also negligible in the wide energy
range, never exceeding 0.01\% 
for the cross-section and 0.1\% 
for the asymmetry.
If an acceptance cut is applied, the changes are yet smaller.

\begin{table}[bht]
\begin{center}
\renewcommand{\arraystretch}{1.1}
\begin{tabular}{|c||c|c|c|c|c|}
\hline
\multicolumn{6}{|c|}{$\afba{\flm}$ with $\theta_{\rm acol}<10^{\circ}$} \\
\hline
$\theta_{\rm acc}$
& $\mz - 3$ & $\mz - 1.8$ & $\mz$ & $\mz + 1.8$ & $\mz + 3$  \\
\hline\hline
$0^{\circ}$ 
      &-0.28487 &-0.16932 & 0.00025 & 0.11500 & 0.16091 \\
      &-0.28453 &-0.16911 & 0.00025 & 0.11486 & 0.16071 \\
\hline
$20^{\circ}$
      &-0.27539 &-0.16367 & 0.00035 & 0.11162 & 0.15635 \\
      &-0.27506 &-0.16347 & 0.00035 & 0.11148 & 0.15616 \\
\hline
$40^{\circ}$
      &-0.24236 &-0.14404 & 0.00050 & 0.09905 & 0.13908 \\
      &-0.24207 &-0.14386 & 0.00050 & 0.09893 & 0.13891 \\
\hline 
\hline
\multicolumn{6}{|c|}{$\afba{\flm}$ with $\theta_{\rm acol}<25^{\circ}$} 
\\
\hline
$\theta_{\rm acc}$
 & $\mz - 3$ & $\mz - 1.8$ & $\mz$ & $\mz + 1.8$ & $\mz + 3$  \\
\hline\hline
$0^{\circ}$ 
      &-0.286732 &-0.170647 &-0.000428 & 0.113029 & 0.156963 \\
      &-0.286474 &-0.170493 &-0.000427 & 0.112927 & 0.156821 \\
\hline
$20^{\circ}$
      &-0.277471 &-0.165114 &-0.000370 & 0.109537 & 0.152173 \\
      &-0.277221 &-0.164965 &-0.000370 & 0.109438 & 0.152036 \\
\hline
$40^{\circ}$
      &-0.244669 &-0.145582 &-0.000255 & 0.096867 & 0.134658 \\
      &-0.244449 &-0.145451 &-0.000255 & 0.096780 & 0.134537 \\
\hline 
\end{tabular}
\caption{\sf
Comparison of \zf\ v.6.11 (first row) with \zf\ v.5.20 (second row)
for the muonic forward-backward asymmetry with
  angular acceptance cut ($\theta_{\rm acc}=0^{\circ},20^{\circ},40^{\circ}$)
  and acollinearity cuts ($\theta_{\rm acol}<10^{\circ}$) and
  ($\theta_{\rm acol}<25^{\circ}$).
The initial-final state interference is switched off and only
final-state radiation is corrected.
\label{tab10fin}
}
\end{center}
\end{table}

\subsection{\large \bf Net corrections
\label{net}
}
Finally, we want to show the resulting effects of the photonic
corrections discussed in the foregoing sections.
We have to distinguish two different approaches to data.
Sometimes experimentalists subtract the intitial-final state interference
contributions 
from measured data, and sometimes the interference effects remain in
the data sample. 

The net corrections without initial-final interferences are negligible 
for the cross-section.
At LEP~1, they are shown for the muonic forward-backward asymmetry 
in Tables \ref{tab10netno}.

\begin{figure}[t] 
\begin{flushleft}
\begin{tabular}{ll}
  \mbox{%
  \epsfig{file=%
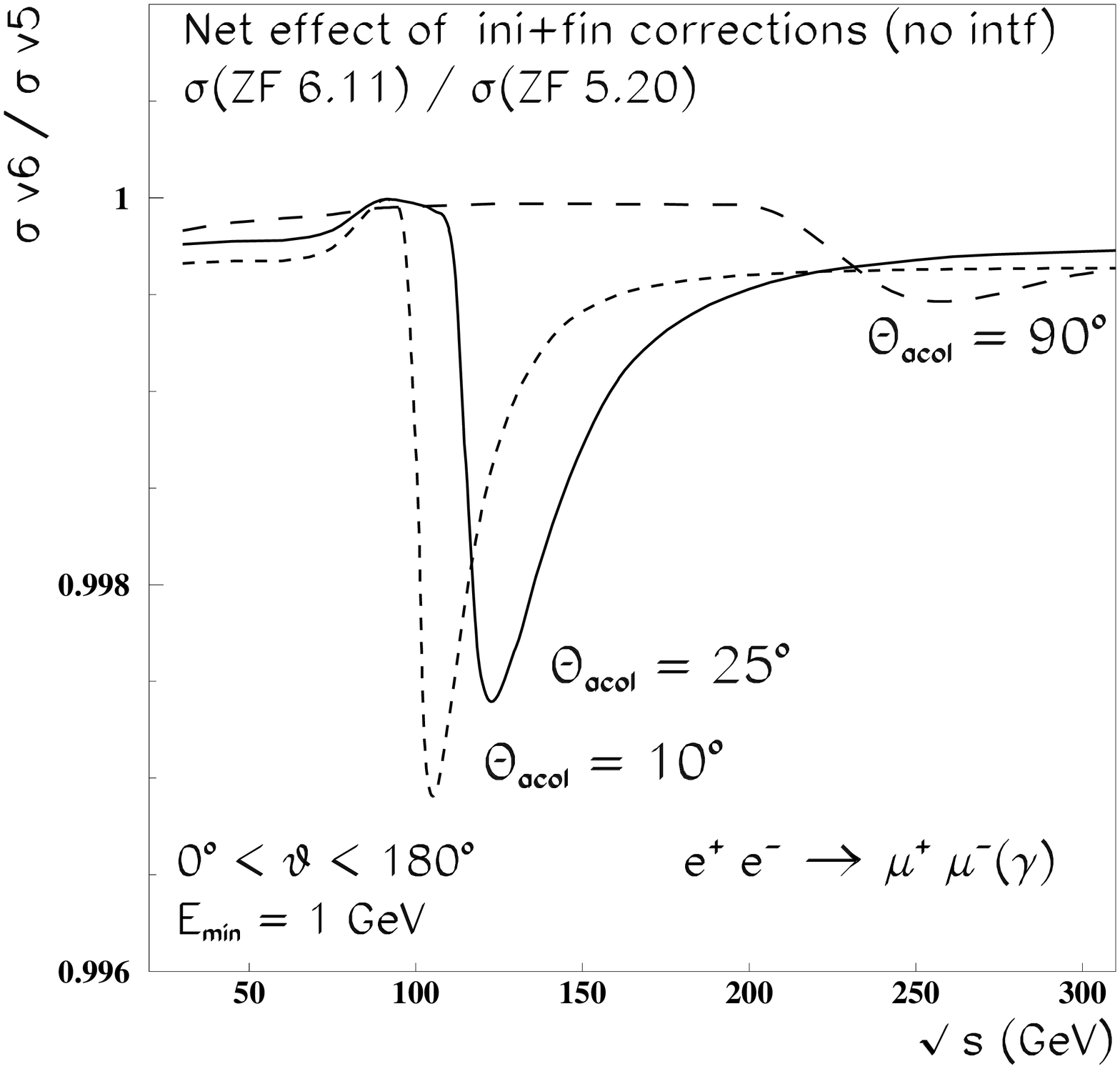
           ,width=7.7cm   
         }}%
&
  \mbox{%
  \epsfig{file=%
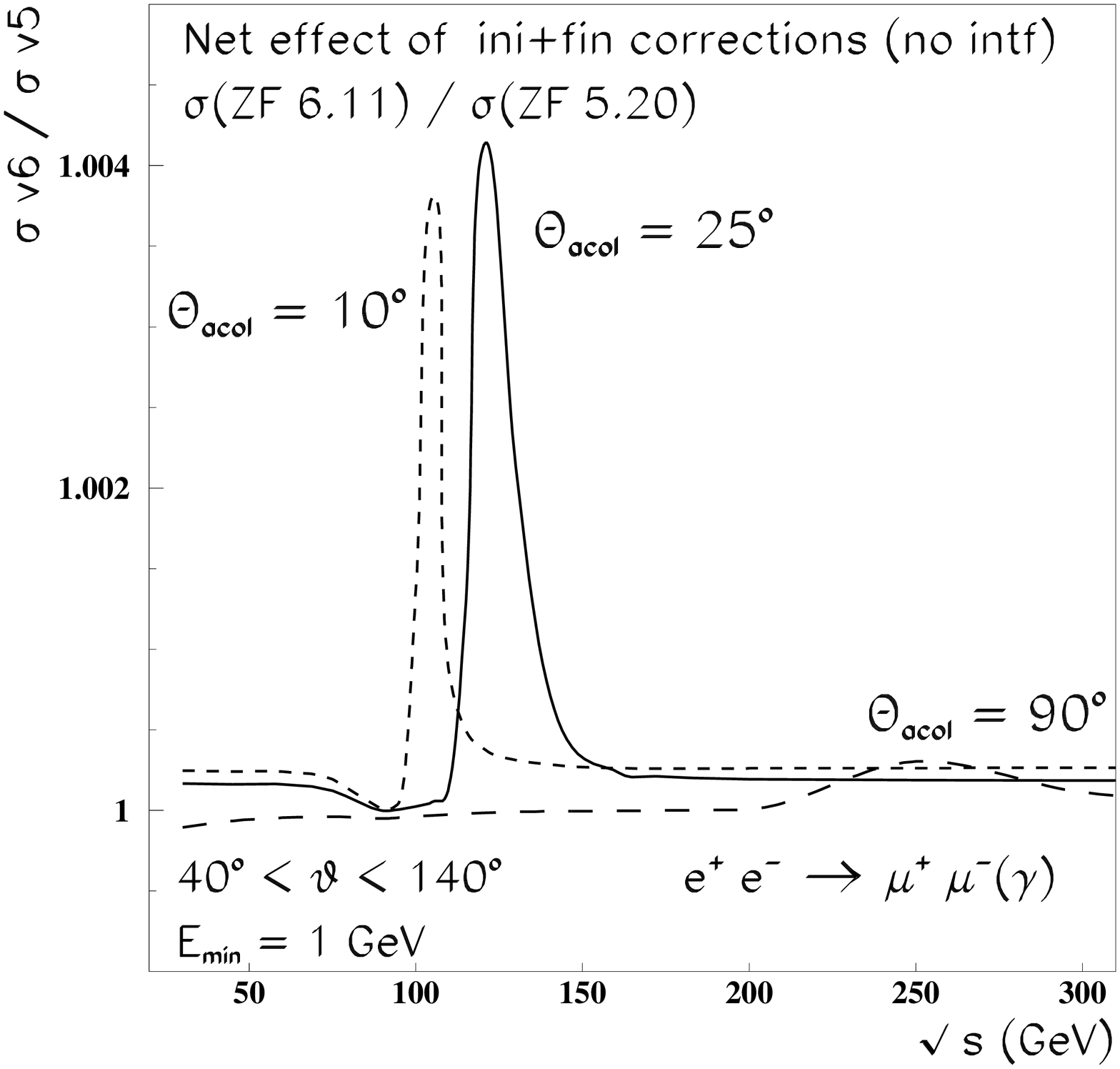
           ,width=7.7cm   
         }}%
\end{tabular}
\caption
{\sf
Net ratios of muon-pair production cross-sections predicted from \zf\ 
v.6.11 and v.5.20 without and with acceptance cut and with three different
acollinearity cuts: $\theta_{\rm acol} < 10^{\circ}, 25^{\circ}, 90^{\circ}$;
$E_{min}=1$ GeV; initial-final state interference not included.
\label{fig26sigc0n} 
}
\end{flushleft}
\end{figure}

{For} a wider energy range, they are shown in Figures \ref{fig26sigc0n}-%
\ref{fig26siga0n}.
Again, at LEP~2 energies, the changes are below what is expected to be
relevant. 

\begin{table}[ht]
\begin{center}
\renewcommand{\arraystretch}{1.1}
\begin{tabular}{|c||c|c|c|c|c|}
\hline
\multicolumn{6}{|c|}{$\afba{\flm}$ with $\theta_{\rm acol}<10^{\circ}$} 
\\
\hline
$\theta_{\rm acc}$
 & $\mz - 3$ & $\mz - 1.8$ & $\mz$ & $\mz + 1.8$ & $\mz + 3$  \\
\hline\hline
$0^{\circ}$ 
      &-0.28497 &-0.16936 & 0.00024 & 0.11496 & 0.16083 \\ 
      &-0.28453 &-0.16911 & 0.00025 & 0.11486 & 0.16071 \\
\hline
$20^{\circ}$
      &-0.27554 &-0.16375 & 0.00032 & 0.11155 & 0.15621 \\
      &-0.27506 &-0.16347 & 0.00035 & 0.11148 & 0.15616 \\
\hline
$40^{\circ}$
      &-0.24259 &-0.14416 & 0.00046 & 0.09893 & 0.13885 \\
      &-0.24207 &-0.14386 & 0.00050 & 0.09893 & 0.13891 \\
\hline 
\hline
\multicolumn{6}{|c|}{$\afba{\flm}$ with $\theta_{\rm acol}<25^{\circ}$} 
\\
\hline
$\theta_{\rm acc}$
 & $\mz - 3$ & $\mz - 1.8$ & $\mz$ & $\mz + 1.8$ & $\mz + 3$  \\
\hline\hline
$0^{\circ}$
      &-0.28677 &-0.17066 &-0.00043 & 0.11302 & 0.15695 \\
      &-0.28647 &-0.17049 &-0.00043 & 0.11293 & 0.15682 \\
\hline
$20^{\circ}$
      &-0.27752 &-0.16514 &-0.00038 & 0.10952 & 0.15214 \\
      &-0.27722 &-0.16497 &-0.00037 & 0.10944 & 0.15204 \\
\hline
$40^{\circ}$
      &-0.24474 &-0.14562 &-0.00027 & 0.09684 & 0.13461 \\
      &-0.24445 &-0.14545 &-0.00026 & 0.09678 & 0.13454 \\
\hline 
\end{tabular}
\caption
{\sf
Comparison of 
net corrections from \zf\ v.6.11 (first row) with \zf\ v.5.20 (second row)
for the muonic forward-backward asymmetry with
  angular acceptance cut ($\theta_{\rm acc}=0^{\circ},20^{\circ},40^{\circ}$)
  and acollinearity cuts ($\theta_{\rm acol}<10^{\circ}$)
  and ($\theta_{\rm acol}<25^{\circ}$).
The initial-final state interference is switched off.
\label{tab10netno}
}
\end{center}
\end{table}

\begin{figure}[t] 
\begin{flushleft}
\begin{tabular}{ll}
  \mbox{%
  \epsfig{file=%
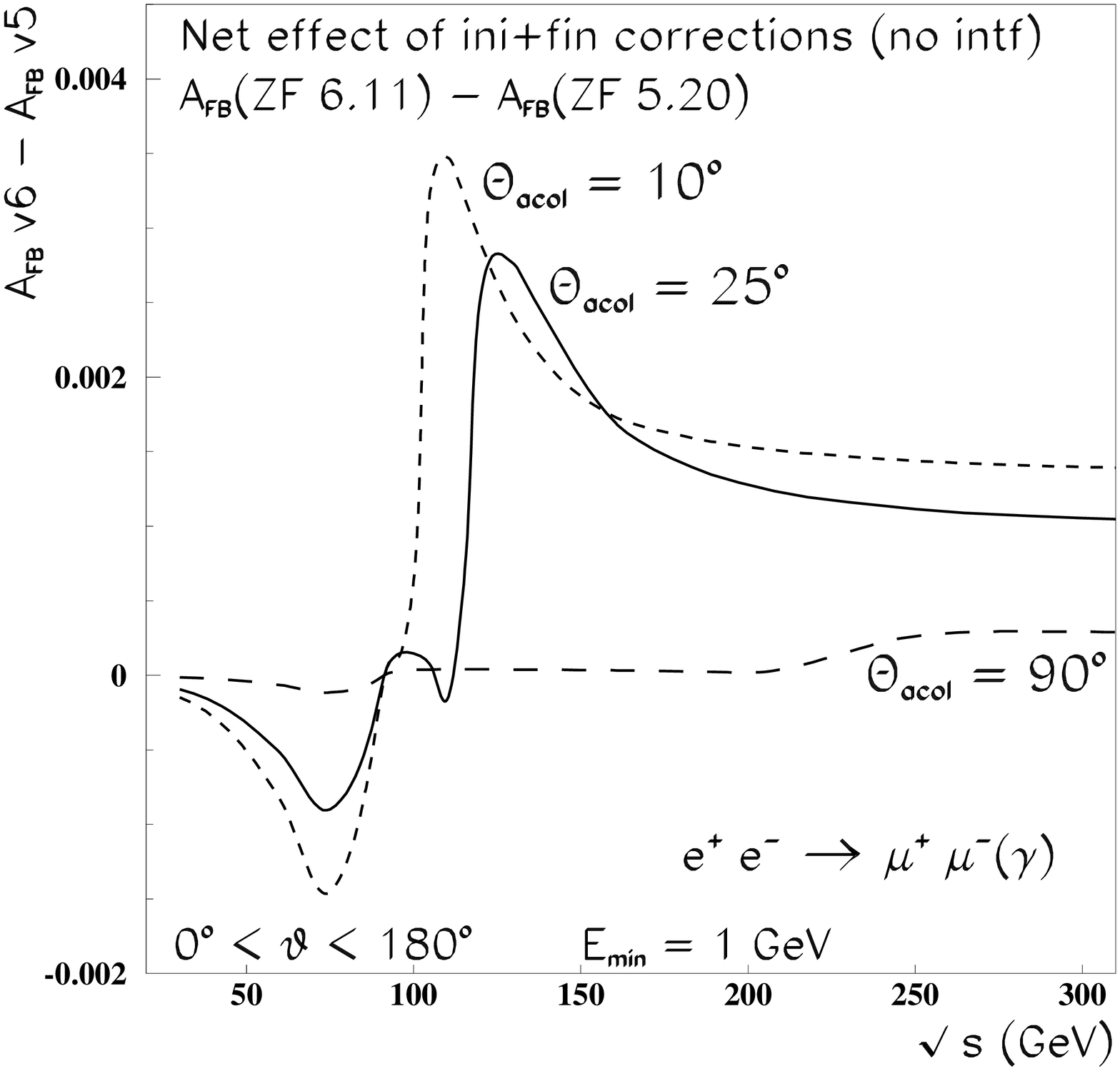
           ,width=7.7cm   
         }}%
&
  \mbox{%
  \epsfig{file=%
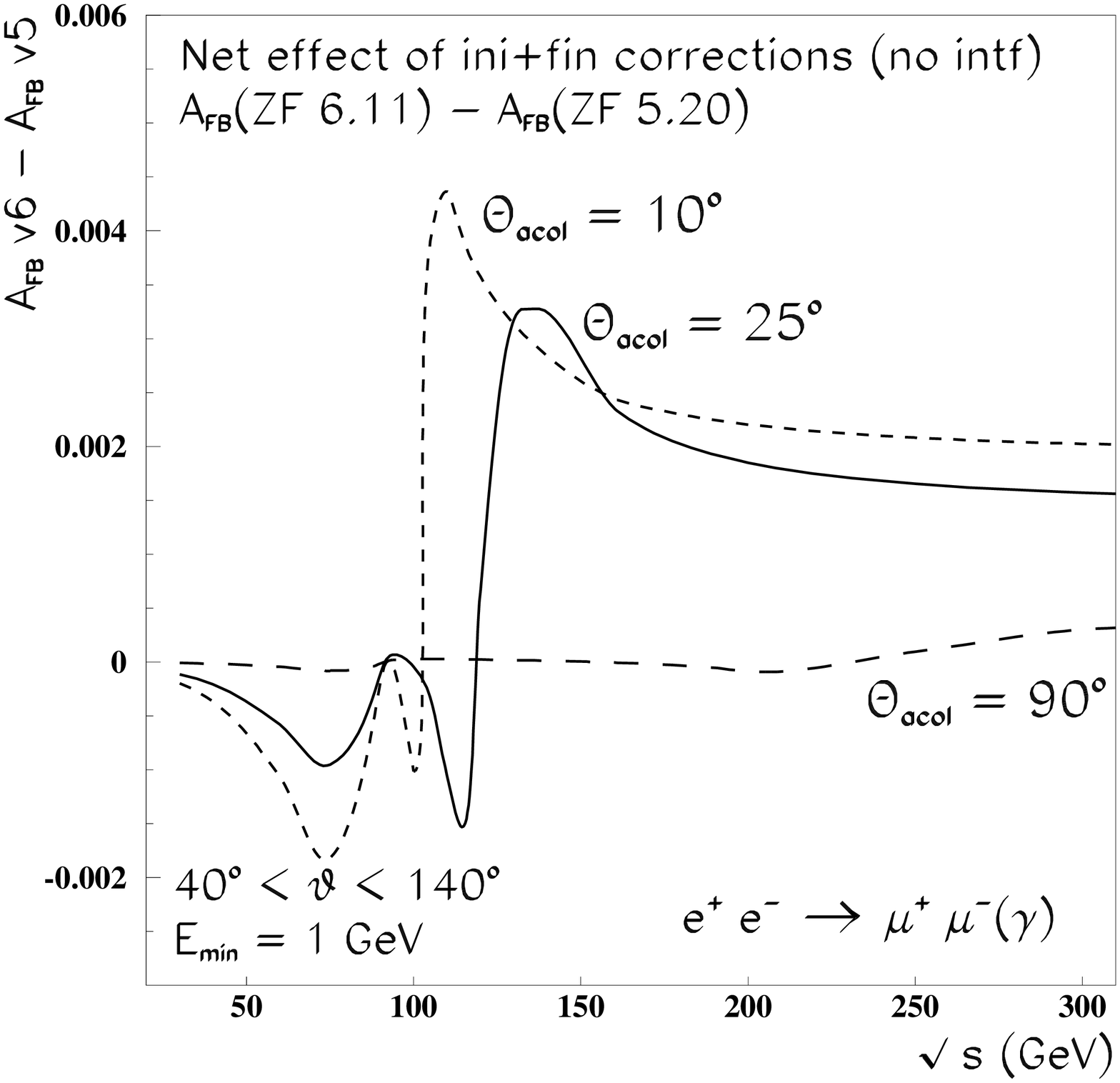
           ,width=7.7cm   
         }}%
\\
\end{tabular}
\caption
{\sf
Net differences of muonic forward-backward asymmetries
predicted from \zf\  
v.6.11 and v.5.20 without and with acceptance cut and with three different
acollinearity cuts: $\theta_{\rm acol} < 10^{\circ}, 25^{\circ}, 90^{\circ}$;
$E_{min}=1$ GeV;  initial-final state interference not included.
\label{fig26afbc0n}
}
\end{flushleft}
\end{figure}
 
We saw that the corrections to the numerical output from \zf\ with
acollinearity cut increased when the corrected initial-final state
interference is taken into account.
The resulting net corrections for the muon production cross-section
and the forward-backward asymmetry at LEP~1 are shown in 
Table \ref{tab10acolc}
and in a wider energy range in Figures
\ref{fig26sigc0n} to \ref{fig26siga0n}.
The numerical effects are dominated by the initial-final state
interference and never exceed 1\% 
at LEP~2 energies.
At LEP~1 they are  much smaller; see the discussion in Section
\ref{interference}.

\begin{table}[ht]
\begin{center}
\renewcommand{\arraystretch}{1.1}
\begin{tabular}{|c||c||c|c|c|c|c|}
\hline
\multicolumn{6}{|c|}{$\sigma_{\flm}\,$[nb] 
with $\theta_{\rm acol}<10^{\circ}$} 
\\
\hline
\hline
$\theta_{\rm acc}$
 & $\mz - 3$ & $\mz - 1.8$ & $\mz$ & $\mz + 1.8$ & $\mz + 3$  \\
\hline\hline
$0^{\circ}$ 
      & 0.21772 & 0.46081 & 1.44776 & 0.67898 & 0.39489 \\
      & 0.21852 & 0.46186 & 1.44782 & 0.67814 & 0.39429 \\
\hline
$20^{\circ}$
      & 0.19869 & 0.42046 & 1.32018 & 0.61877 & 0.35972 \\
      & 0.19892 & 0.42075 & 1.32021 & 0.61857 & 0.35959 \\
\hline
$40^{\circ}$
      & 0.14974 & 0.31675 & 0.99349 & 0.46515 & 0.27019 \\
      & 0.14978 & 0.31680 & 0.99350 & 0.46511 & 0.27016 \\
\hline\hline
\multicolumn{6}{|c|}{$\afba{\flm}$ with $\theta_{\rm acol}<10^{\circ}$} 
\\
\hline
$0^{\circ}$ 
      &-0.28222 &-0.16710 & 0.00083 & 0.11392 & 0.15926 \\ 
      &-0.28282 &-0.16783 & 0.00070 & 0.11475 & 0.16059 \\
\hline
$20^{\circ}$
      &-0.27319 &-0.16187 & 0.00080 & 0.11066 & 0.15486 \\
      &-0.27408 &-0.16261 & 0.00070 & 0.11133 & 0.15594 \\
\hline
$40^{\circ}$
      &-0.24093 &-0.14294 & 0.00074 & 0.09837 & 0.13797 \\
      &-0.24151 &-0.14343 & 0.00069 & 0.09890 & 0.13888 \\
\hline 
\hline
\multicolumn{6}{|c|}{$\sigma_{\flm}\,$[nb] 
 with $\theta_{\rm acol}<25^{\circ}$} 
\\
\hline
 & $\mz - 3$ & $\mz - 1.8$ & $\mz$ & $\mz + 1.8$ & $\mz + 3$  \\
\hline\hline
$0^{\circ}$ 
      & 0.22228 & 0.46836 & 1.46602 & 0.68816 & 0.40127 \\
      & 0.22281 & 0.46905 & 1.46603 & 0.68754 & 0.40081 \\
\hline
$20^{\circ}$
      & 0.20281 & 0.42728 & 1.33688 & 0.62731 & 0.36571 \\
      & 0.20321 & 0.42781 & 1.33689 & 0.62684 & 0.36536 \\
\hline
$40^{\circ}$
      & 0.15280 & 0.32183 & 1.00618 & 0.47188 & 0.27502 \\
      & 0.15287 & 0.32192 & 1.00619 & 0.47180 & 0.27496 \\
\hline\hline
\multicolumn{6}{|c|}{$\afba{\flm}$ with $\theta_{\rm acol}<25^{\circ}$} 
\\
\hline
$0^{\circ}$ 
      &-0.28580 &-0.16975 &-0.00000 & 0.11296 & 0.15683 \\
      &-0.28555 &-0.16975 &-0.00005 & 0.11307 & 0.15701 \\
\hline
$20^{\circ}$
      &-0.27684 &-0.16451 &-0.00006 & 0.10952 & 0.15213 \\
      &-0.27657 &-0.16447 &-0.00009 & 0.10963 & 0.15229 \\
\hline
$40^{\circ}$
      &-0.24445 &-0.14540 &-0.00010 & 0.09696 & 0.13476 \\
      &-0.24444 &-0.14542 &-0.00011 & 0.09700 & 0.13483 \\
\hline 
\end{tabular}
\caption{\sf
Comparison of 
net corrections from 
\zf\ v.6.11 (first row) with \zf\ v.5.20 (second row)
for muon-pair production with
  angular acceptance cut ($\theta_{\rm acc}=0^{\circ},20^{\circ},40^{\circ}$)
  and acollinearity cuts ($\theta_{\rm acol}<10^{\circ}$)
  and ($\theta_{\rm acol}<25^{\circ}$).
The initial-final state interference is switched on.
\label{tab10acolc}
}
\end{center}
\end{table}

\begin{figure}[t] 
\begin{flushleft}
\begin{tabular}{ll}
  \mbox{%
  \epsfig{file=%
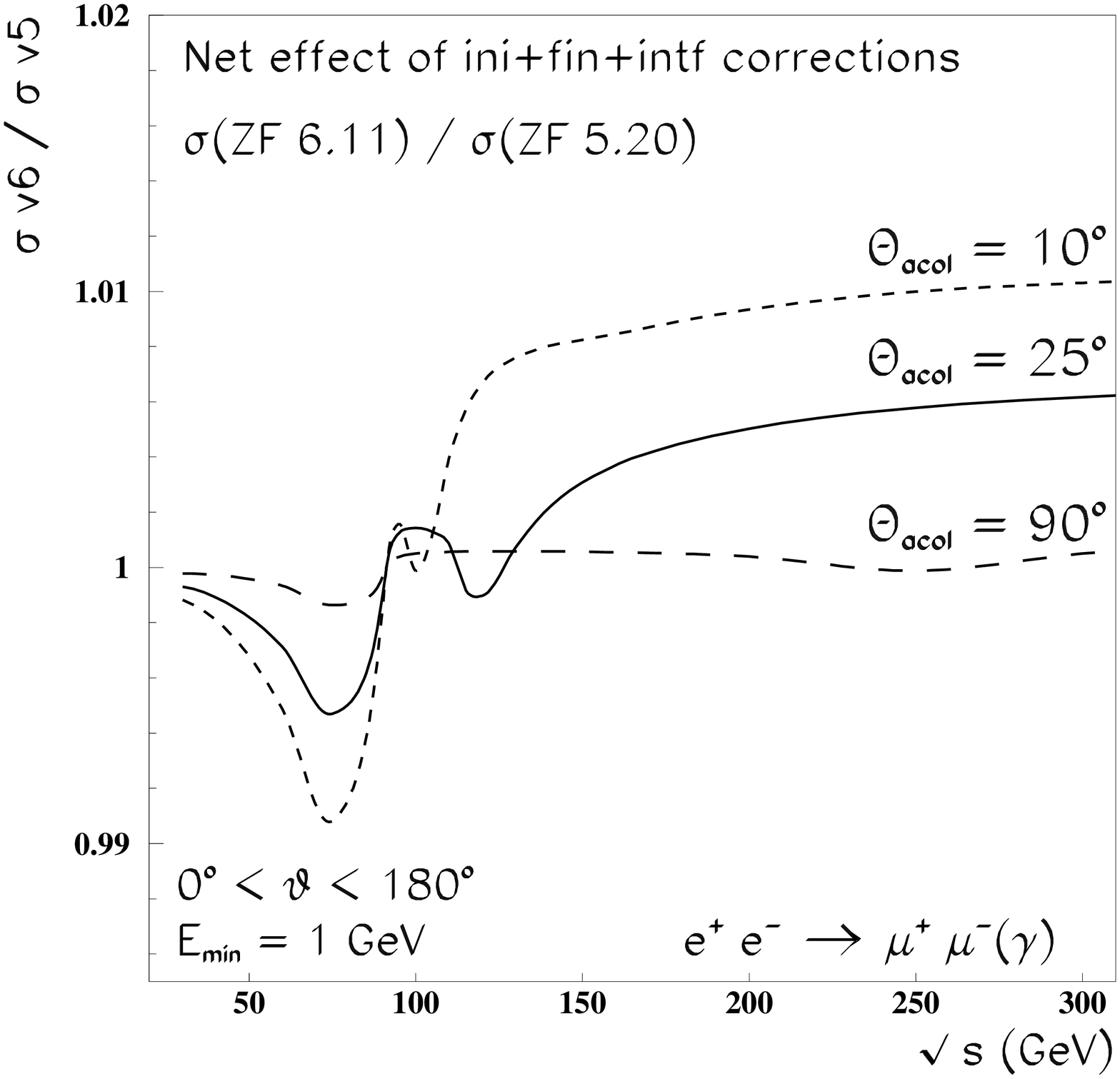
          ,width=7.7cm   
           }}%
&
  \mbox{%
  \epsfig{file=%
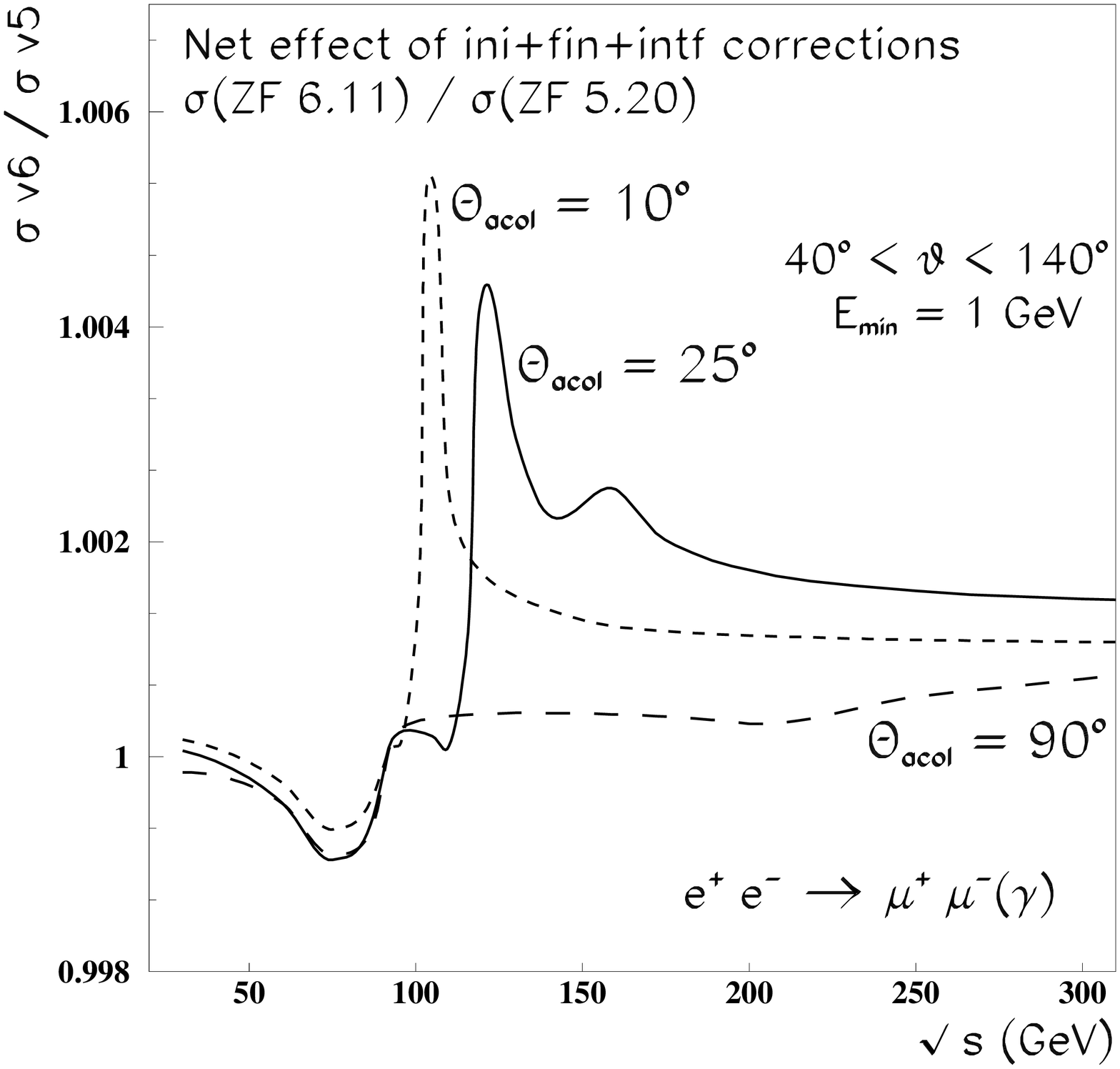
           ,width=7.7cm   
           }}%
\\
\end{tabular}
\caption
{\sf
Net ratios of muon-pair production cross-sections predicted from \zf\ 
v.6.11 and v.5.20 without and with acceptance cut and with three different
acollinearity cuts: $\theta_{\rm acol} < 10^{\circ}, 25^{\circ}, 90^{\circ}$;
$E_{min}=1$ GeV; all the discussed corrections included.
\label{fig26siga0n}
}
\end{flushleft}
\end{figure}

\begin{figure}[t] 
\begin{flushleft}
\begin{tabular}{ll}
  \mbox{%
\epsfig{file=%
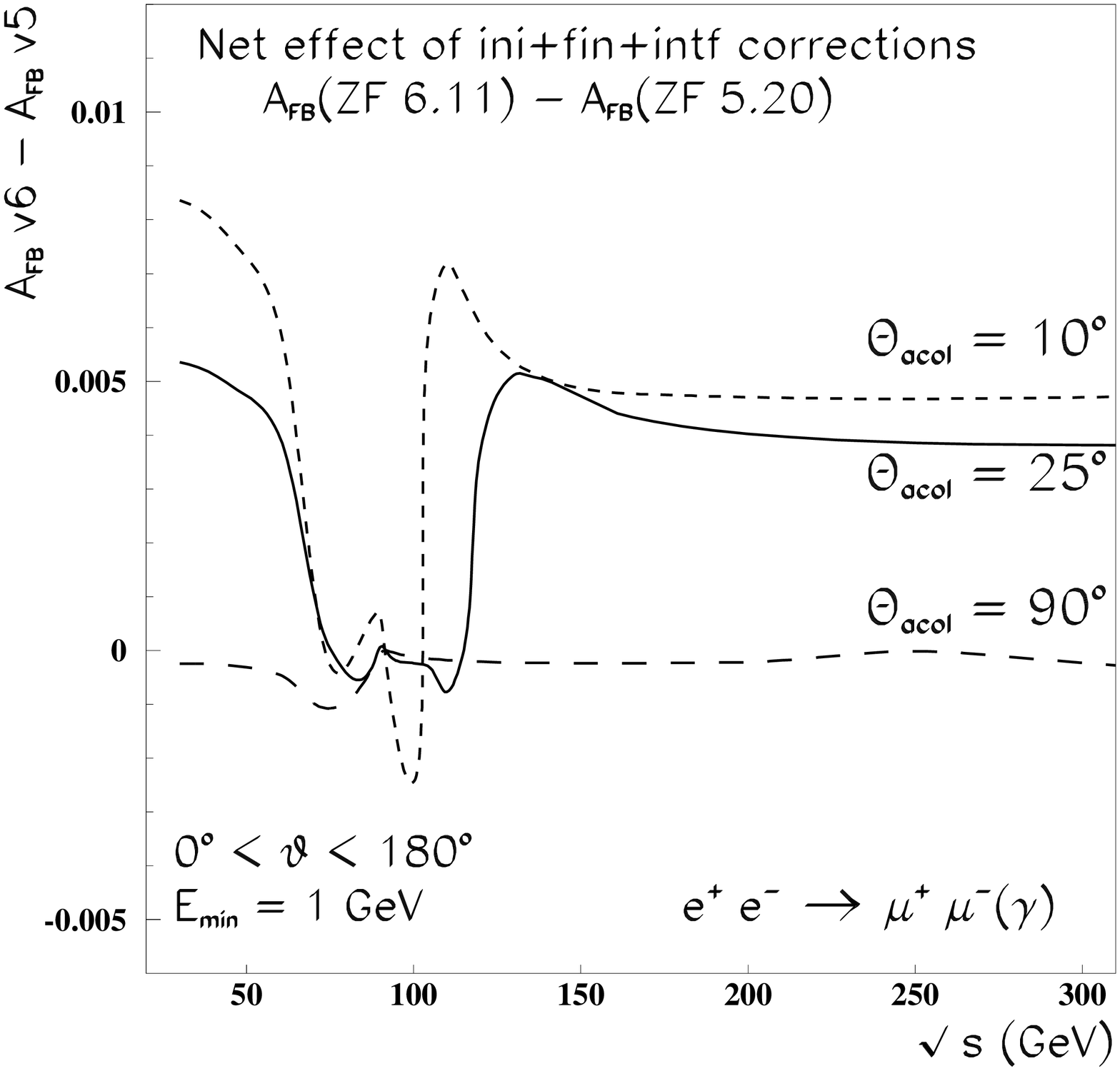
           ,width=7.7cm   
         }}%
&
  \mbox{%
\epsfig{file=%
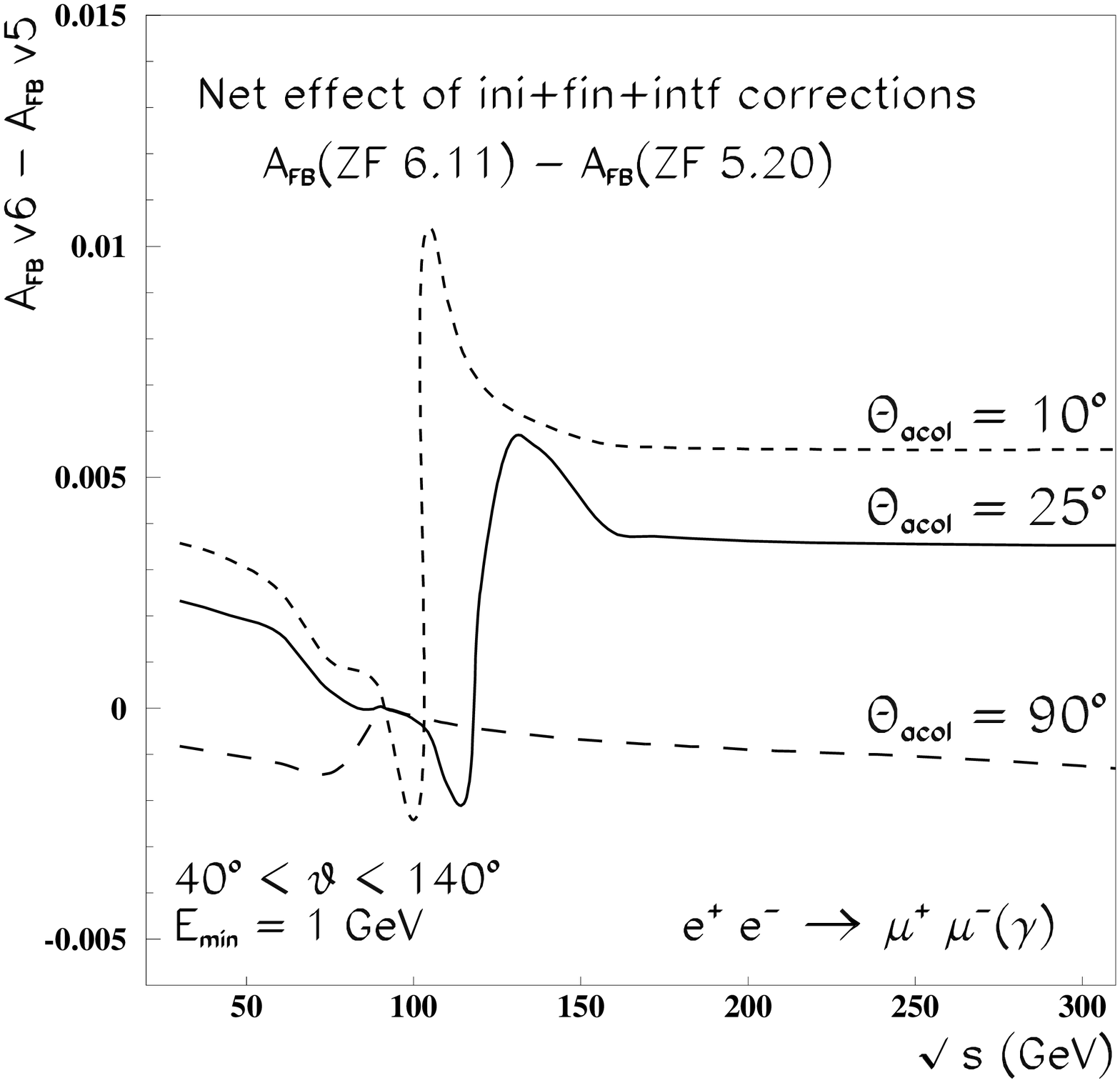
           ,width=7.7cm   
         }}%
\\
\end{tabular}
\caption
{\sf
Net differences of muon-pair production asymmetries predicted from \zf\ 
v.6.11 and v.5.20 without and with acceptance cut and with three different
acollinearity cuts: $\theta_{\rm acol} < 10^{\circ}, 25^{\circ}, 90^{\circ}$;
$E_{min}=1$ GeV; all the discussed corrections included.
\label{fig26afbac0n}
}
\end{flushleft}
\end{figure}

\section{\Large \bf Some comparisons with other programs
\label{summary}
}
We conclude this report on the update of \zf\ with few comparisons
with other programs .

For LEP~1, we restrict ourselves to the two small tables in Table
\ref{tab10acolifi2}.
They represent an update of Tables 37 and 38 of \cite{Bardin:1999gt},
which were produced with \zf\ v.5.20.
The complete Tables 37--40 (the latter two for another acollinearity
cut) may be updated with use of our Tables
\ref{tab10sigifi} and \ref{tab10afbifi}.
The agreement was not considered to be satisfactory in
\cite{Bardin:1999gt} but is excellent now. 

\begin{table}[bhtp]
\begin{center}
\renewcommand{\arraystretch}{1.1}
\begin{tabular}{|c||c|c|c|c|c|}
\hline
\multicolumn{6}{|c|}{
{$\sigma_{\flm}\,$[nb] with $\theta_{\rm acol}<10^\circ$}}
\\ 
\hline
$\theta_{\rm acc} = 0^\circ$& $\mz - 3$ & $\mz - 1.8$ & $\mz$ & $\mz + 1.8$ &
$\mz + 3$  \\ 
\hline\hline
  & 0.21932  & 0.46287  & 1.44795  & 0.67725  & 0.39366 \\
{{\tt TOPAZ0}}  & 0.21776  & 0.46083  & 1.44785  & 0.67894  & 0.39491 \\
  & 
{\bf --7.16}    & 
{\bf --4.43}     &
{\bf --0.07}     &
{\bf +2.49}     &
{\bf +3.17}    
\\ 
\hline
  & 0.21928  & 0.46284  & 1.44780  & 0.67721  & 0.39360 \\
{{\tt ZFITTER}}  & 0.21772  & 0.46082  & 1.44776  & 0.67898  &
0.39489 \\ 
  &
{\bf --7.16}     &{\bf --4.40}     &{\bf --0.03 }    &{\bf +2.60}
&{\bf +3.27}   
\\ 
\hline 
\hline
\multicolumn{6}{|c|}{$\afba{\flm}$ with $\theta_{\rm acol}<10^\circ$} \\
\hline
$\theta_{\rm acc} = 0^\circ $& $\mz - 3$ & $\mz - 1.8$ & $\mz$ & $\mz + 1.8$ &
$\mz + 3$  \\ 
\hline\hline
  & --0.28450 & --0.16914  & 0.00033  & 0.11512  & 0.16107 \\
{{\tt TOPAZ0}}  & --0.28158 & --0.16665  & 0.00088  & 0.11385  &
0.15936 \\ 
  & 
{\bf +2.92}    & {\bf +2.49}     &
{{\bf +0.55}}     &{\bf --1.27}
&
{{\bf --1.71}}    \\ 
\hline
  & --0.28497 & --0.16936  & 0.00024  & 0.11496  & 0.16083 \\
{{\tt ZFITTER}}  & --0.28222 & --0.16710  & 0.00083  & 0.11392  & 0.15926 \\
  & {\bf +2.75}    & {\bf +2.27}     & {\bf +0.60}    &{\bf --1.03}
&{\bf --1.56}
\\
\hline 
\end{tabular}
\caption[]{
{\sf
A comparison of predictions for muonic cross-sections and forward-backward
asymmetries around the $Z$ peak.
First row is without initial-final state interference, second row with,
third row the relative effect of that interference in per mil.
}
}
\label{tab10acolifi2}
\end{center}
\end{table}

\begin{figure}[t] 
\begin{flushleft}
\begin{tabular}{ll}
  \mbox{%
  \epsfig{file=%
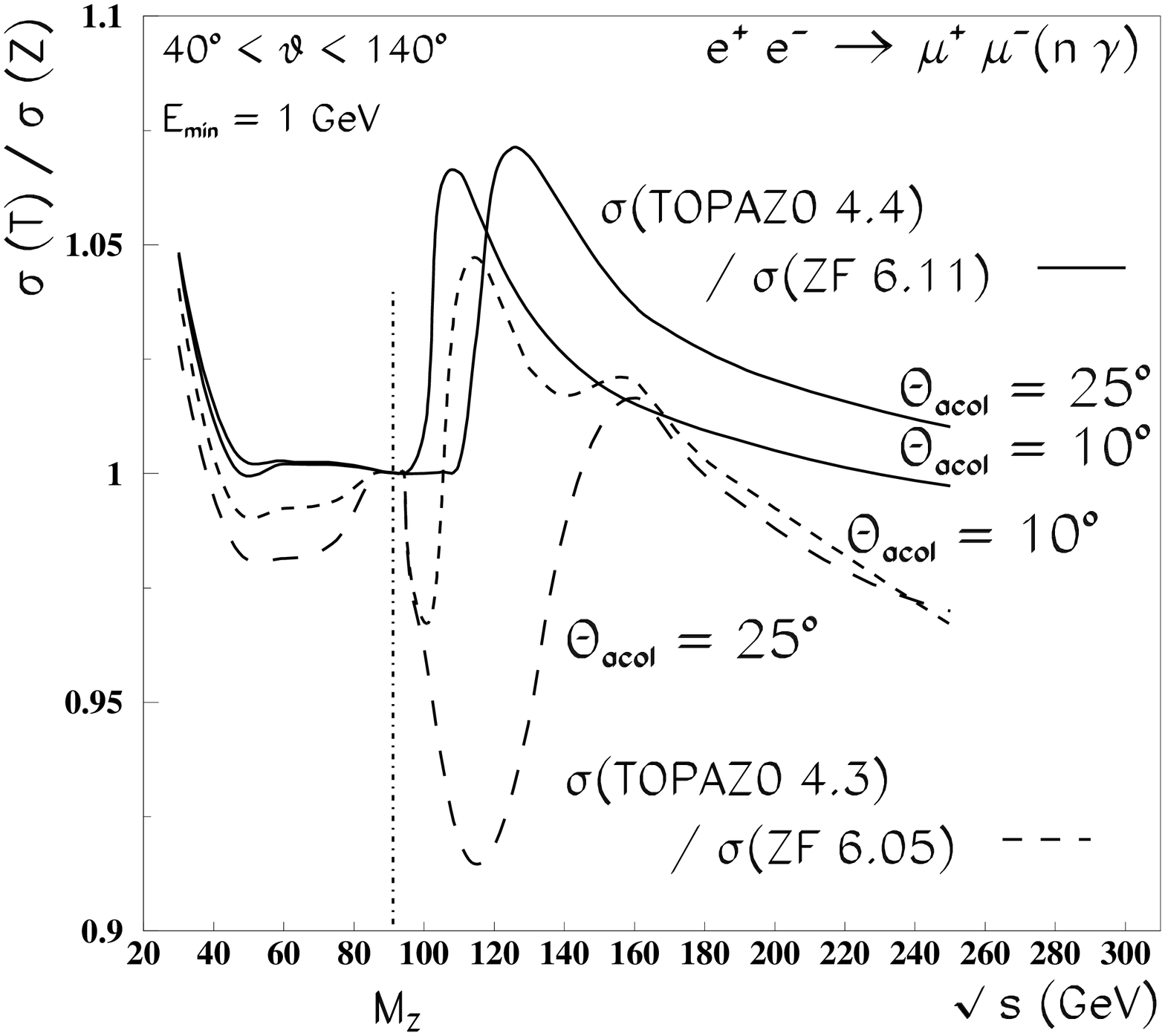
           ,width=8.cm   
         }}%
&
  \mbox{%
  \epsfig{file=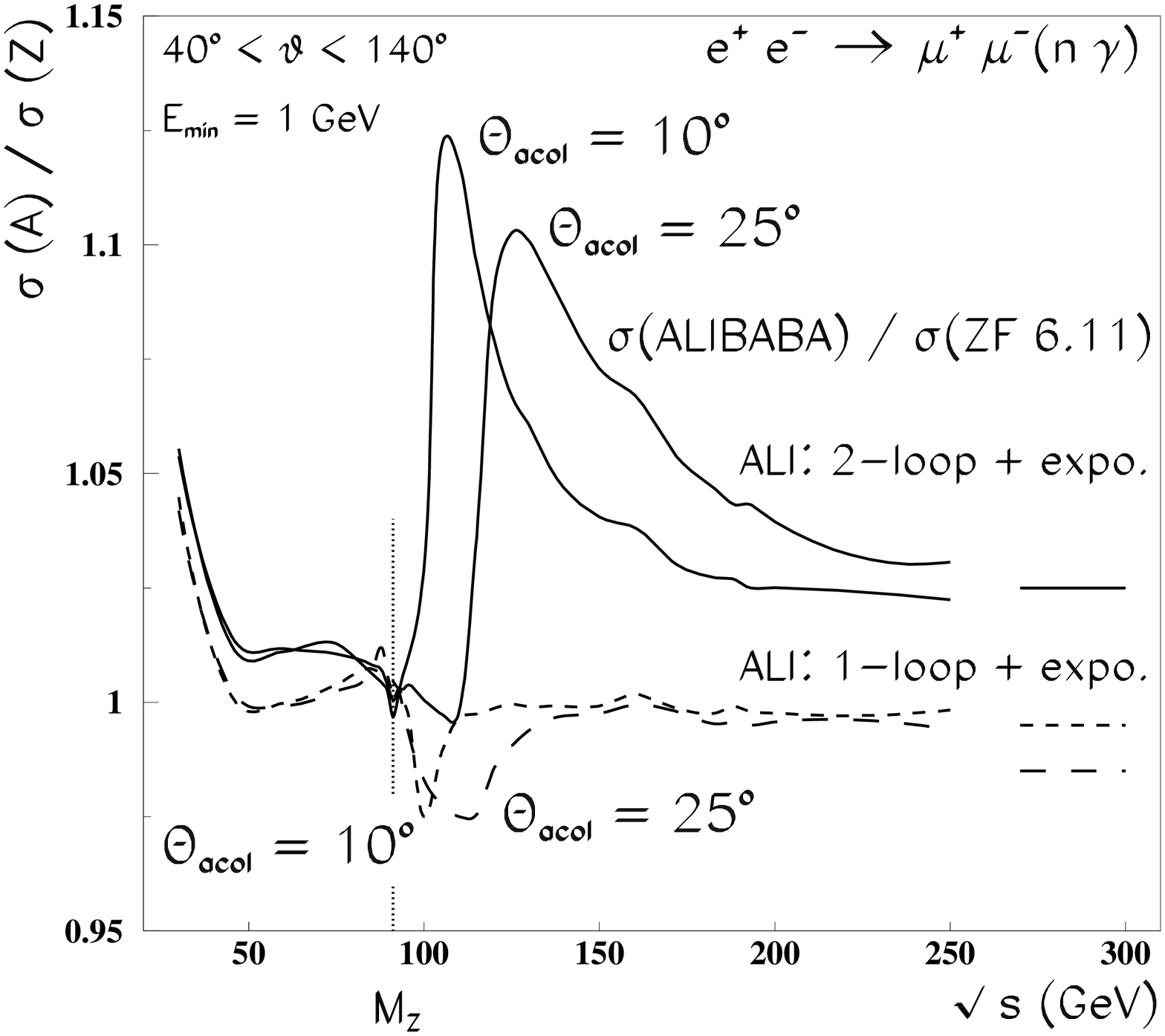,width=8.cm}}
\\
\end{tabular}
\caption
{\sf
Muon-pair production cross-section ratios with 
$\theta_{\rm acol}^{\max}=10^{\circ}, 25^{\circ}$ and 
$\theta_{\rm acc}=40^{\circ}$;
(a) {\tt TOPAZ0} v.4.3 and v.4.4 versus {\tt ZFITTER} v.6.04/06 and
v.6.11 (1999),  (b) {\tt ALIBABA} v.2 (1990) versus {\tt ZFITTER} v.6.11.
Flag setting: {\tt ISPP}=0.
\label{compar-tzaz}
}
\end{flushleft}
\end{figure}

For a wider energy range, we performed a first comparison in Figure 3
of \cite{Christova:1998tc}, based on {\tt ALIBABA} v.2 (1990)
\cite{Beenakker:1991mb},   
{\tt TOPAZ0} v.4.3 (1999)
\cite{Montagna:1998kp,Bardin:1999gt,Passarino:199800}, and 
{\tt ZFITTER} v.5.14 (1998). 
We have repeated the comparison with the same version of  {\tt
ALIBABA}, {\tt TOPAZ0} v.4.4 (1999) \cite{Montagna:1998kp,topaz0}, and
{\tt ZFITTER} v.6.11 (1999) in Figure \ref{compar-tzaz}.
The \zf\ numbers are produced with the default settings (if not
otherwise stated), and the other two programs have also been run by
ourselves.

Concerning the {\tt TOPAZO} v.4.4 ratios, we register a different behaviour
(compared to v.4.3) for 
$\theta_{\rm acc}=40^{\circ}$ which is now much closer to the
{\tt ALIBABA} ratios.
In \zf\, we varied the treatment of some higher-order corrections via
flags {\tt FOT2} and {\tt PAIRS} with not too much effect.
Furthermore, when we switched off  the two-loop
contributions in {\tt ALIBABA} (with setting {\tt IORDER=3}), the
agreement became much better.

\begin{figure}[t] 
\begin{flushleft}
\begin{tabular}{ll}
  \mbox{%
  \epsfig{file=%
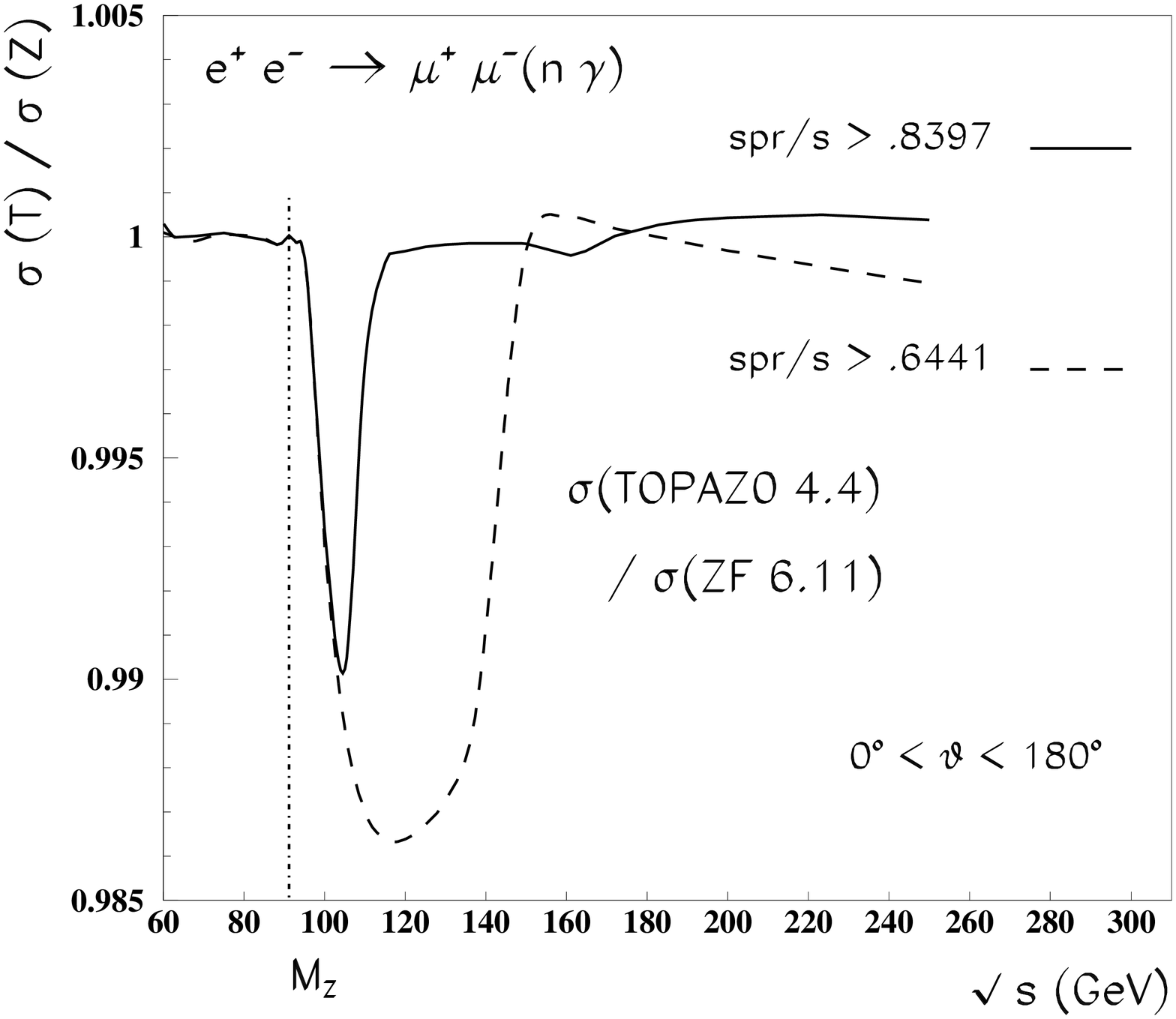
,width=8.cm   
         }}%
&
  \mbox{%
  \epsfig{file=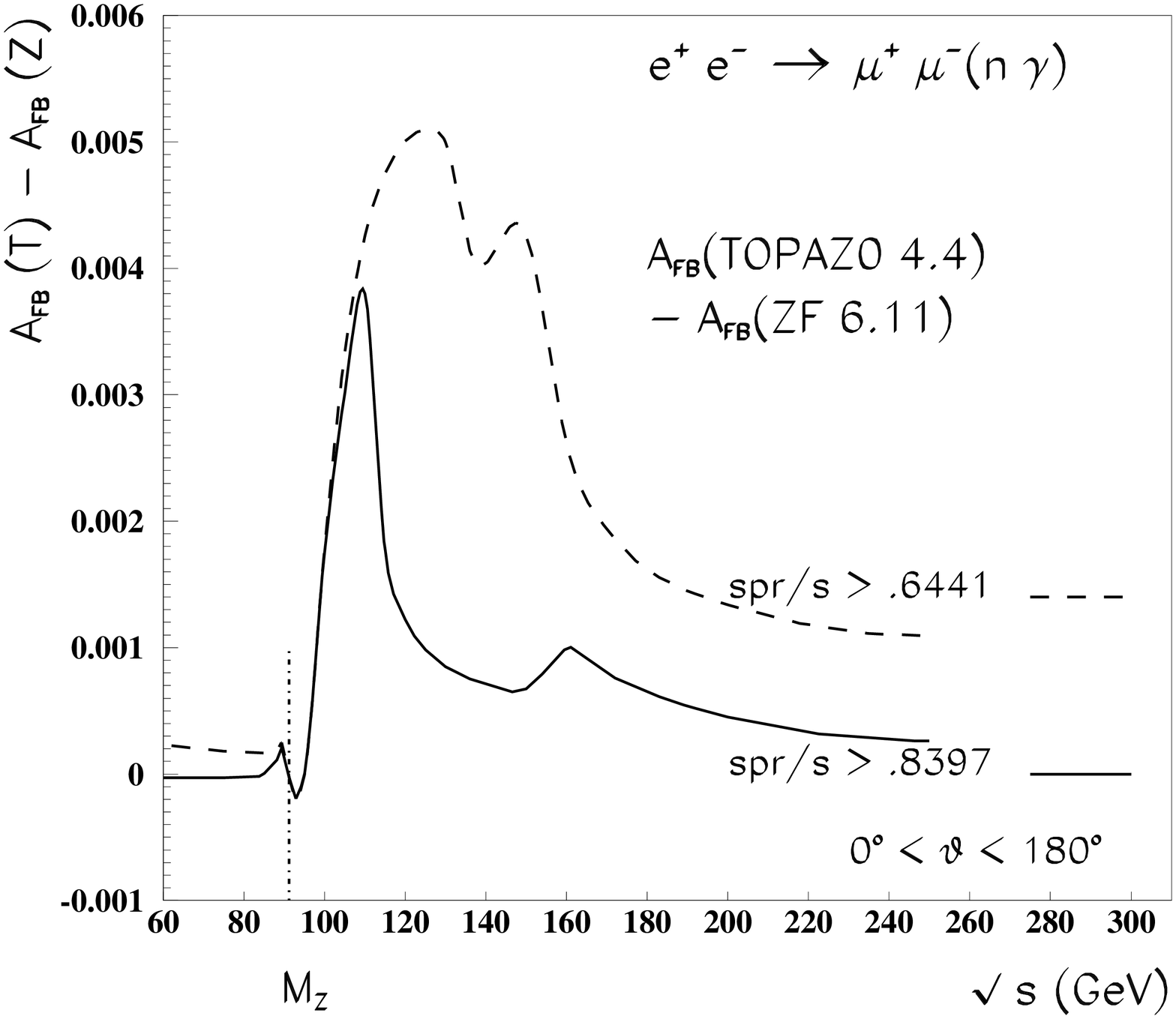,width=8.cm}}
\\
\end{tabular}
\caption[]
{\sf
Comparison of predictions from  {\tt ZFITTER} v.6.11  and 
{\tt TOPAZ0} v.4.4 with $s'$-cut: (a) 
Muon-pair production cross-section ratios, (b) forward-backward
asymmetry differences.
Flag setting: {\tt ISPP}=1, {\tt FINR}=0; further: 
{\tt SIPP}={\tt S\_PR} \cite{DESY99070}. 
\label{compar-spr}}
\end{flushleft}
\end{figure}

On the other hand, a cross check of the {\tt ZFITTER} and {\tt TOPAZ0} 
programs applying $s'$-cuts comparable to the acollinearity cuts
used for the figures above and below show a very high level of 
agreement between the two, at LEP~1 ($< O(3\cdot 10^{-4})$), 
but also at LEP~2 energies at the order of less than a per mil
\footnote{For LEP~2 energies flag 
{\tt FINR} was set to 0 for the final-state corrections -- recommended 
choice.} Initial-state pair production and exponentiation of higher-orders 
do not spoil this high level of agreement for the $s'$-cut.
This may be seen in Figures \ref{compar-spr}. 

Our $s'$-cut dependent ratios deviate from unity mostly in the regions 
where the radiative return is not prevented.
The same is true for the ratios with acollinearity cut; since this cut is
not as effective in preventing the radiative return as the $s'$-cut,
the deviations survive at higher energies to some extent.
This fact and the higher order corrections, which remained untouched
by our study, seem to be the main sources of the remaining deviations
between the different programs.

In this context, it will be quite interesting to see the comparisons
with $s'$-cut 
of \zf\ and {\tt KORALZ} \cite{Jadach:1994yv} and {\tt KK}
\cite{Jadach:1998jb} 
being extended to situations  with acollinearity cut (see also
\cite{home-Jadach,Jadach:1999gz}).   

\newpage
\section{\Large \bf Summary
\label{summary2}
}
We derived analytical formulae for the photonic corrections with
acollinearity cut and got substantial deviations from the coding in \zf\
until version 5.
The essentials of the changes have been described and numerical
comparisons are performed in great detail. 
Fortunately, we may conclude that the numerical changes are not as big as
one could expect.
Although, certain differences to predictions of other codes remain
untouched. 
They are pronounced when the radiative return to the $Z$ resonance is
kinematically allowed.
If one is interested to perform investigations in this kinematical regime
further studies are needed. 

\clearpage

\section*{References}

\def\href#1#2{#2} 

 \clearpage
\def\thesection{\Alph{section}}
\setcounter{section}{0}
\appendix
\def\theequation{\Alph{section}.\arabic{equation}}
\def\thetable{\Alph{section}.\arabic{table}}
\def\thefigure{\Alph{section}.\arabic{figure}}
\setcounter{equation}{0}
\setcounter{table}{0}
\setcounter{figure}{0}
\section{\Large \bf Radiative Return and Acollinearity Cut
\label{app-ret}
}
\newcommand{\nc}{\newcommand}
\nc{\GeV}{\,\mbox{GeV}}
\nc{\MeV}{\,\mbox{MeV}}
\nc{\pb}{\,\mbox{pb}}
\nc{\nb}{\,\mbox{nb}}
\nc{\btu}{\bigtriangleup}
\nc{\DD}{\displaystyle}
An acollinearity cut may act as a simple cut on invariant masses and
thus it may prevent the radiative return 
of $\sqrt{s'}$ to the $Z$ peak (and the development of the radiative tail) 
for measurements at higher $\sqrt{s}$.
In Figure \ref{dalitz} it may be seen that a reasonable analogue of a cut
value $\sqrt{s^{min}}$ is the upper value of $s'$ of region III,
$R_{\theta_{acol}}$, 
defined in (\ref{eq:rx}):
\ba
\nonumber
{\DD\sqrt{s^{\min}}}
&=&
 \frac{M_Z}{\sqrt{R_{\theta_{acol}}}},
\ea

with

\ba
\nonumber
{\DD R_{\theta_{acol}} }
&=& 
\frac{1-\sin{({\DD
{\theta_{\rm acol}^{\max}}}/{\DD 2})}}
{1+\sin{({\DD {\theta_{\rm acol}^{\max}}}/{\DD 2})}}.
\ea

The relations are visualised in Table \ref{rxivalues}.
 
\begin{table}[ht]
\begin{center}
\begin{displaymath}
\begin{array}{|r@{.}l|r@{.}l|r@{.}l|} 
\hline
\multicolumn{2}{|c|}{}&\multicolumn{2}{c|}{}&\multicolumn{2}{c|}{}
\\
 \multicolumn{2}{|c|}{\raisebox{1.5ex}{${\DD \xi^{\max}}$}} 
& \multicolumn{2}{c|}{\raisebox{1.5ex}[-1.5ex]
{
$
{\DD R_\xi 
}
$
}} 
& \multicolumn{2}{c|}{\raisebox{1.5ex}
{
${{\DD\sqrt{s^{\min}}
}}
$
}}
\\\hline\hline
     ~~~0&0^\circ~   & ~1&0000~   & ~~~~91&2~\GeV~~
\\\hline
     2&0^\circ   & 0&9657   & 92&8~\GeV
\\\hline
     5&0^\circ   & 0&9164   & 95&3~\GeV
\\\hline
    10&0^\circ   & 0&8397   & 99&5~\GeV
\\\hline
    15&0^\circ   & 0&7691  & 104&0~\GeV
\\\hline
    20&0^\circ   & 0&7041  & 108&7~\GeV
\\\hline
    25&0^\circ   & 0&6441  & 113&6~\GeV
\\\hline
    30&0^\circ   & 0&5888  & 118&8~\GeV
\\\hline
    45&0^\circ   & 0&4465  & 136&5~\GeV
\\\hline
    60&0^\circ   & 0&3333  & 157&9~\GeV
\\\hline
    75&0^\circ   & 0&2432  & 184&9~\GeV
\\\hline
    90&0^\circ   & 0&1716  & 220&1~\GeV
\\\hline
   120&0^\circ   & 0&0718  & 340&3~\GeV
\\\hline
   150&0^\circ   & 0&0173  & 692&6~\GeV
\\\hline
   180&0^\circ   & 0&     &\multicolumn{2}{c|}{\infty}
\\\hline
\end{array}
\end{displaymath}
\caption
{\sf
The 
minimal cms energy, $\sqrt{s^{\min}}$,
 at which the radiative return
to the $Z$ peak is prevented by an acollinearity cut given as a
function of this cut.
\label{rxivalues}
}
\end{center}
\end{table}

\end{document}